\documentclass{emulateapj}

\usepackage{subfigure}
\usepackage{graphics,graphicx}
\usepackage{lscape}

\newcommand{\chan}{{\it Chandra}}

\newcommand{\kms}{$\,\rm{km\,s^{-1}}$}
\newcommand{\msolar}{$\rm{M}_{\odot}$}
\newcommand{\rxj}{RXJ1720.1+2638}
\newcommand{\bulletclus}{1ES0657-558}
\shorttitle{Cold Front Clusters from the {\it Chandra} Archive}
\shortauthors{Owers et al.}
\begin{document}
\title{A High Fidelity Sample of Cold Front Clusters from the {\it CHANDRA} Archive}
\author{Matt S. Owers\altaffilmark{1,3}, Paul E.J. Nulsen\altaffilmark{2}, 
Warrick J. Couch\altaffilmark{3}, Maxim Markevitch\altaffilmark{2}}
\altaffiltext{1}{School of Physics, University of New South Wales, Sydney, NSW 
2052, Australia; mowers@astro.swin.edu.au}
\altaffiltext{2}{Harvard Smithsonian Center for Astrophysics, 60 Garden Street,
 Cambridge, MA 02138, USA}
\altaffiltext{3}{Center for Astrophysics and Supercomputing, Swinburne 
University of Technology, Hawthorn, VIC 3122, Australia}

\begin{abstract}
This paper presents a sample of ``cold front'' clusters selected from the \chan\
archive. The clusters are selected based purely on the existence of  
surface brightness edges in their \chan\ images which are modeled as density 
jumps. A combination of the derived density and temperature jumps 
across the fronts is used to select nine robust examples of cold front 
clusters: \bulletclus, Abell~1201, Abell~1758N, MS1455.0+2232, Abell~2069, 
Abell~2142, Abell~2163, \rxj, and Abell~3667. This sample is the subject of an 
ongoing study aimed at relating cold fronts to cluster merger activity, and 
understanding how the merging environment affects the cluster constituents. 
Here, temperature maps are presented along with the \chan\ X-ray images. A 
dichotomy is found in the sample in that there exists a subsample of cold front 
clusters which are clearly mergers based on their X-ray morphologies, and a 
second subsample which harbor cold fronts, but have surprisingly relaxed X-ray 
morphologies, and minimal evidence for merger activity at other wavelengths. 
For this second subsample, the existence of a cold front provides the sole 
evidence for merger activity at X-ray wavelengths. We discuss how cold fronts
can provide additional information which may be used to constrain merger 
histories, and also the possibility of using cold fronts to distinguish
major and minor mergers.

\end{abstract}

\keywords{galaxies: clusters: individual: 1ES0657-558, Abell~1758N, 
MS1455.0+2232,  Abell~2069, Abell~2142, Abell~2163, RXJ1720.1+2638, Abell~3667,
 Abell~665, Abell~2034 
 --- X-rays: galaxies: clusters }

\section{Introduction}
The hierarchical nature of structure formation is well understood theoretically 
\citep{press1974,lacey1993,peebles1993,springel2006} and well supported 
observationally \citep{smoot1992,hernquist1996,cole2005,jeltema2005}. At the
present epoch, clusters of galaxies are the largest and most massive  
large-scale structures that are virialized and they are growing 
hierarchically through the gradual accretion of surrounding matter. Occasionally,
two clusters of roughly equal mass fall together in a major merger, releasing 
around $10^{64}$\,erg of gravitational potential energy during the merger 
process \citep{markevitch1999}. To a degree, the majority of clusters exhibit 
evidence for merger activity. However it is the {\it scale} of the cluster 
mergers, i.e., the distinction between major mergers and the continuing minor 
mergers/infall of smaller subsystems, that is difficult to quantify. A gauge of 
merger activity capable of distinguishing major and minor mergers would greatly 
assist to understand mergers and their effects on clusters and their constituents.

A solution may be provided by high resolution X-ray observations with the \chan\
observatory \citep{weisskopf2002}. Among its first discoveries was the detection 
and characterization of the ``cold front'' phenomenon in the intracluster media
(ICM) of Abell~2142 \citep{markevitch2000} and Abell~3667 
\citep{vikhlinin2001b}. Cold fronts reveal themselves as edge-like features in 
the \chan\ images and are well modeled as contact discontinuities which form at
the interface of cool, dense gas and hotter, more diffuse gas. Simulations 
show cold fronts can arise in more than one way during cluster mergers 
\citep{poole2006, ascasibar2006}. Broadly speaking, cold fronts can be separated
into two general types: ``remnant core'' and ``sloshing'' \citep[for a review 
see][]{markevitch2007}. 

Remnant core type cold fronts were first invoked to explain the origin of the 
cold fronts in Abell~2142 \citep{markevitch2000}. In this case, the 
discontinuity occurs between the interface of an infalling subcluster's cool 
core and the hotter ambient ICM of the main cluster. The subcluster's core is 
bared to the hotter ICM after ram pressure stripping removes its 
less dense, lower pressure outer atmosphere during its passage through the 
main cluster. The cool core survives longest because, at first, it remains bound
to the infalling dark matter core and, after separating from that, because of its
high density. This explanation for Abell~2142's cold fronts has since been 
abandoned in favor of a ``sloshing'' type mechanism \citep{markevitch2007}. 
An excellent example of remnant core type cold front is observed in the 
Bullet cluster \citep[\bulletclus;][]{markevitch2002} while a similar mechanism 
acts to form cold fronts in the elliptical galaxies NGC~1404 and NGC~4552 
\citep[where the cold fronts are at the interface of the galaxy's gas and the 
cluster ICM;][]{machacek2005,machacek2006}. Furthermore, remnant core type cold 
fronts have also been seen in both idealized \citep{ascasibar2006,poole2006,
springel2007,mastropietro2008} and cosmological \citep{bialek2002,nagai2003,
onuora2003,mathis2005} simulations of cluster mergers.

Gas ``sloshing'' came to the fore as an explanation for the existence of a cold 
front in the ``relaxed'' cluster Abell~1795 \citep{markevitch2001}. These cold
fronts occur when the stably stratified ICM is disturbed in such a way that the 
cool, dense gas residing in the cluster core is displaced from its position
at the bottom of the gravitational potential well. The cold front is the contact
discontinuity formed at the interface of the low entropy core gas and the higher
entropy gas at larger radii. In most postulated scenarios, the disturbance is 
gravitational in nature with its origin in an infalling subcluster 
\citep[e.g.][]{tittley2005,ascasibar2006}. However, there have been suggestions 
the disturbance may be a hydrodynamic phenomenon in the form of weak shocks or 
acoustic waves \citep{churazov2003,fujita2004}, while \citet{markevitch2003asp} 
suggest AGN explosions may be responsible, although the example of Hydra~A 
used was later shown to be better interpreted as a weak shock 
\citep{nulsen2005}. To date, the \citet{ascasibar2006} study provides the most
comprehensive set of simulations for the sloshing scenario and show it is 
possible that the cold fronts observed in otherwise relaxed appearing clusters 
(based on their X-ray morphologies) can be produced by sloshing induced in 
relatively minor mergers. Currently, there is no observational evidence that 
the cold fronts in these relaxed appearing clusters are related to merger 
activity. 

While cold fronts offer interesting insights into the physical properties of 
the ICM \citep[e.g., ][]{vikhlinin2001a,churazov2004,lyutikov2006,
markevitch2007,xiang2007}, it is how their presence relates to cluster merger 
activity, along with the tantalizing possibility of utilizing cold fronts as 
a gauge for cluster mergers, which provides the motivation for the study 
presented in this paper. {\it The utilization of cold fronts as merger gauges 
becomes even more compelling when considering those clusters which appear to 
harbor relaxed X-ray morphologies---the existence of a cold front provides the 
only indication of ongoing merger activity at X-ray wavelengths}. Much work has 
gone into studying cold front formation 
with simulations, but what has been lacking to date is a systematic study of the 
relationship between cold fronts and other dynamical indicators of cluster 
merger activity. The number of clusters observed by \chan\, to harbor cold 
fronts has grown significantly and now allows a representative sample to be 
assembled in order to perform such a study. To this end, this paper focuses on 
the selection of a sample of clusters with cold fronts in 
their \chan\, X-ray images. The selection criteria are designed to ensure that 
the high quality X-ray data shows an unambiguous cold front in each member of the
sample.

To date, our goal of establishing whether there is a link between the presence of
cold fronts in clusters and evidence of recent dynamical growth has been met, at
least partially, in two cases, Abell~1201 \citep{owers2009a} and Abell~3667 
\citep{owers2009b}. In these studies, significant dynamical substructure was 
detected using spatial and redshift information provided by large samples of 
spectroscopically confirmed cluster member galaxies. In these clusters, the 
presence of a cold front is directly related to the presence of substructures, 
and thus ongoing merger activity. Future work will endeavor to go one step 
further and study the galaxy properties within the cold front clusters. 

The outline of this paper is as follows. In Section~\ref{selcrit} the procedure 
for selecting cold front clusters is presented. In Section~\ref{tempmaps} the 
technique used to generate projected temperature maps for each cold front cluster
is described. In Section~\ref{cf.clusters} the \chan\, X-ray images and 
temperature maps for the clusters in the cold front sample are presented, along 
with a brief description of the X-ray properties and evidence for merger activity
from the literature. In Section~\ref{discussion}, the results and sample are discussed.
In Section~\ref{summary} summary and conclusions are presented.

\section{Sample selection criteria}\label{selcrit}
\subsection{Initial Selection}
To ensure selection of only the best defined cold front clusters, the following
criteria were set. The data must be of high enough quality that 
statistically robust measurements of density and temperature can be made on 
either side of the edge in surface brightness for characterization. The 
bulk of the cluster X-ray emission must be captured on the \chan\, chips, 
whilst resolution and cosmological surface brightness dimming affects must be 
minimized. Clearly, the cluster must also exhibit an edge in the surface 
brightness, indicating the existence of a density discontinuity caused by a cold
front. 

The initial sample was selected from those clusters with publicly available data
within the \chan\, archive (as of July 2006) and the preceding conditions were 
imposed as the following quantitative criteria:

\begin{itemize}
\item a total \chan\ ACIS-I and/or ACIS-S exposure time exceeding 40\,ks;
\item cluster redshift in the range $0.05\leq z\leq 0.3$; 
\item a significant edge in the X-ray surface brightness.
\end{itemize}

The density discontinuities caused by cold fronts are generally 
quite visible in the \chan\ images, and their existence is easily verified by 
the observation of edges and compression of isophotal contours overlaid 
onto the images. Thus, for each \chan\ data set meeting the first three 
selection criteria, the 0.5 -- 7\,keV band image was scrutinized for edge-like 
features and regions of compressed isophotes. It is noted that the
morphology of the X-ray emission was not considered during the selection 
process, nor were any clusters selected based on prior knowledge of a cold front
in their ICM. The significance of the density discontinuity was quantified by 
modeling the surface brightness across the edge with a broken power law density 
model (see Appendix~\ref{sb.model}). The surface brightness edge must also be
reasonably sharp, i.e., in order to be classified as a front, they must be 
reasonably well fitted by the density jump model. 

This model was used to further cull the 
sample and is presented in Appendix~\ref{sb.model}. A description of the method 
used to fit the model to the data follows.

\subsection{Fitting the surface brightness across the edges}\label{sb.fit}
Corresponding to the broken power law density model, the surface brightness 
model given in Equation~\ref{sb.mod} was incorporated 
into the {\it Sherpa} fitting software, which is part of the {\it Chandra 
Interactive Analysis of Observations} ({\it CIAO}; version 3.4 is used here) 
package. Point sources were identified using the {\it CIAO} WAVDETECT tool and 
removed for fitting.  The images were restricted to the energy range 
0.5 -- 7\,keV, so that background and instrument effects were minimized, 
whilst still allowing enough source photons to obtain good fits to the models. 
Where possible, separate observations had their data and exposure maps co-added 
prior to fitting. The background count rates for \chan\ observations is known to
change with time, and also to differ between the ACIS-S and ACIS-I arrays. For 
these reasons, observations taken at significantly different times, or taken 
using different chips were not co-added, but were simultaneously
fitted with separate background components. The fit was restricted to regions 
where there are edges (the blue regions shown in the top left panels of 
Figures~\ref{fig:1e657a} through \ref{fig:a2142a}), and the initial inputs for 
the center of 
the spheroid, the ellipticity, $\epsilon$, the orientation angle, $\theta$, and 
the radius at which the edge occurs, $R_f$, were obtained by overlaying an 
elliptical region that best approximated the surface brightness contours across 
the edge. The center of curvature was not necessarily the position of 
the X-ray centroid of the cluster, and the semimajor axis of the ellipse, 
$\theta$, was oriented to bisect the edge. Since the center, $\theta$ 
and $R_f$ are highly degenerate, the center and $\theta$ were fixed to those 
values determined from the manually fitted ellipse and only $R_f$ was determined
directly from the fit. A fixed, uniform background component was added to the 
model during fitting. The background was determined by fitting the entire 
\chan\, observed area using one or two beta models plus a constant component. 
The beta model/s account for the cluster emission, and the constant component 
for the background. We note that this constant background component does not 
correctly account for the unvignetted instrumental background. This does not 
affect our results because: a) the edges are found in regions where the cluster 
emission is dominant and b) the quantity of interest is the break in the surface
brightness profile which is insensitive to both the background level and to small 
changes in the background slope (when the background does not dominate). 
Each model was multiplied by the exposure map, generated using the {\it CIAO} 
MKINSTMAP, ASPHIST and MKEXPMAP tools, which account for instrumental effects 
such as vignetting, Quantum Efficiency (QE), QE nonuniformity, bad pixels, 
dithering and effective area.

\begin{figure*}
  \begin{center}
      {\includegraphics[angle=-90,width=0.31\textwidth]{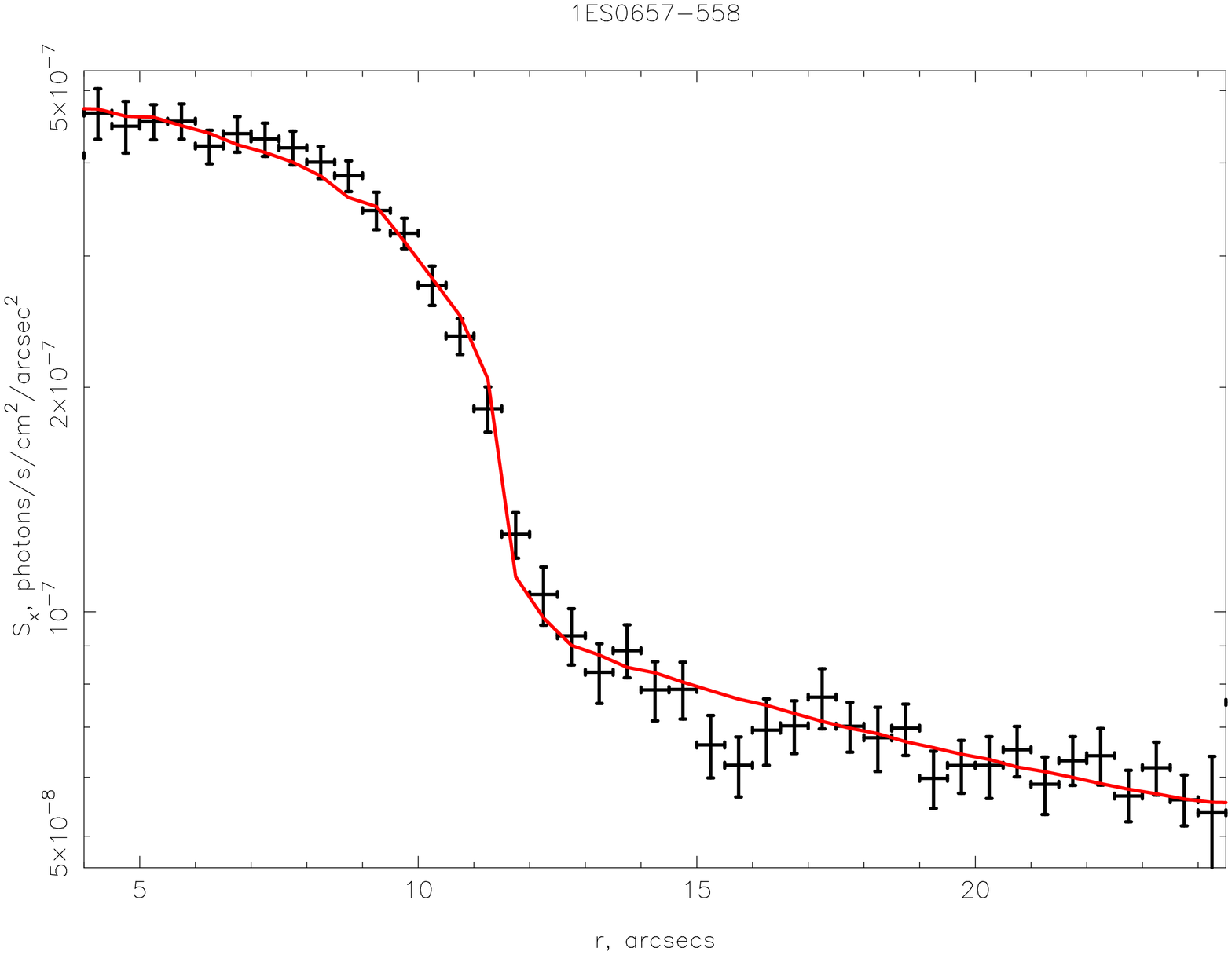}}
      {\includegraphics[angle=-90,width=0.31\textwidth]{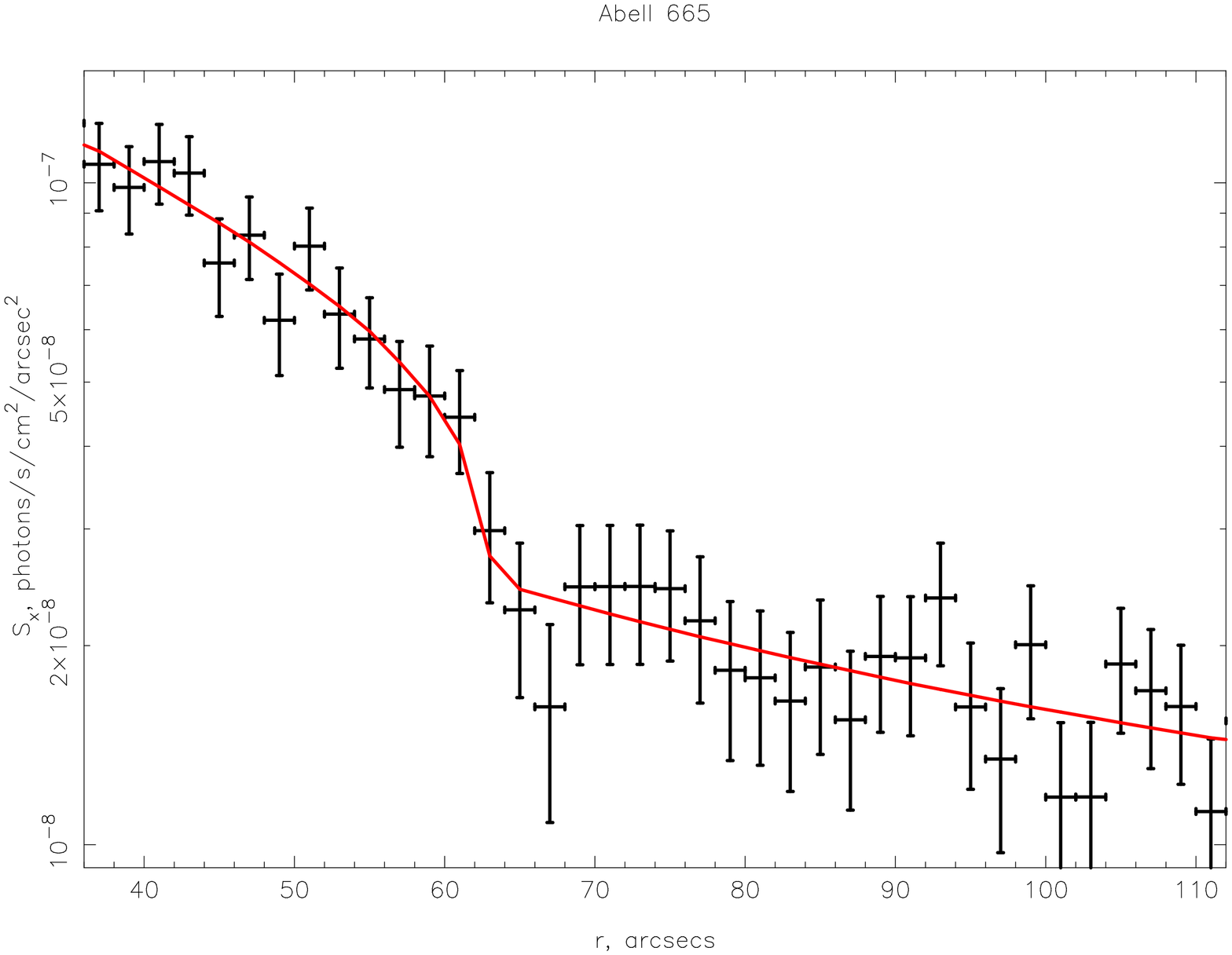}}
      {\includegraphics[angle=-90,width=0.31\textwidth]{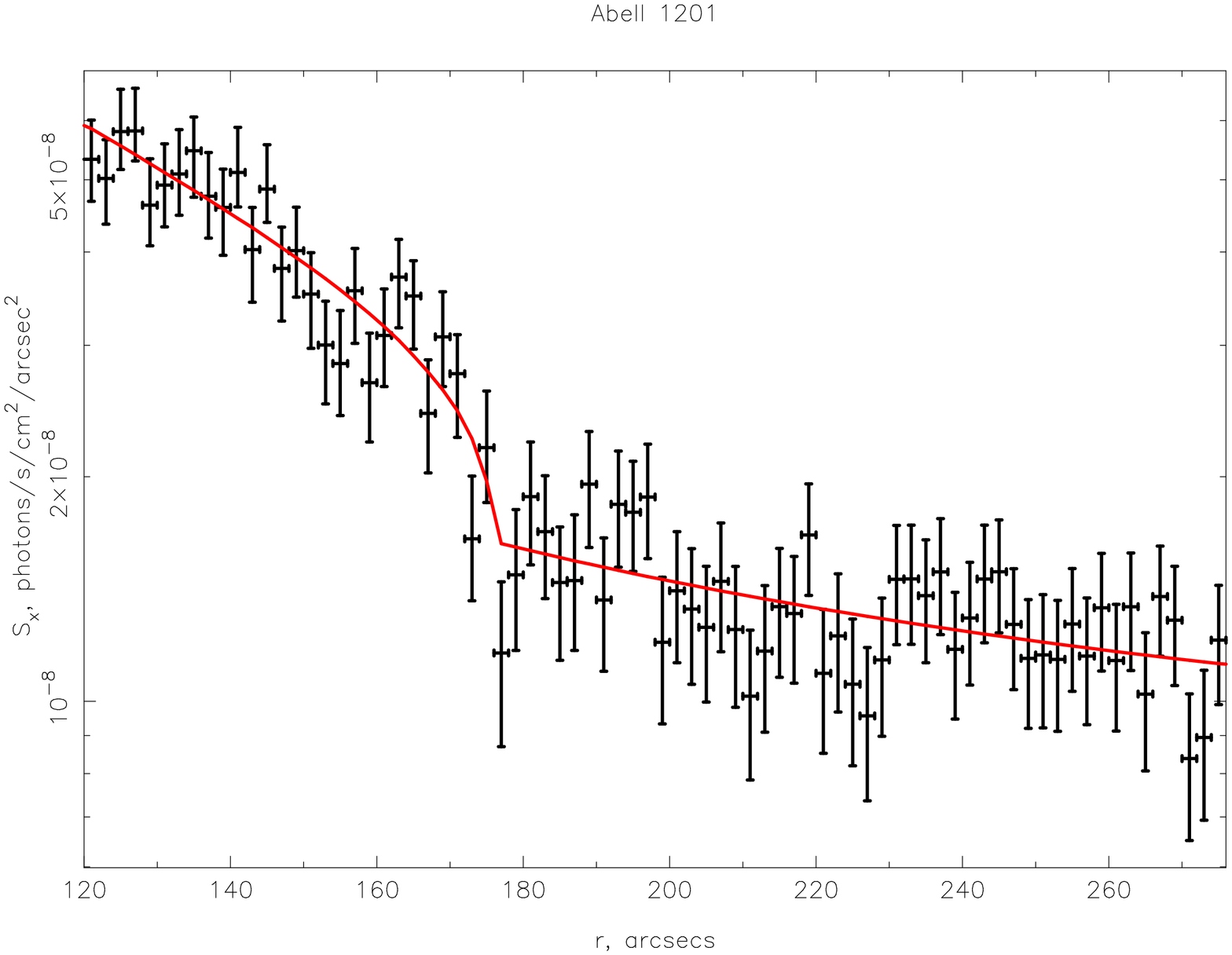}}\\
      {\includegraphics[angle=-90,width=0.31\textwidth]{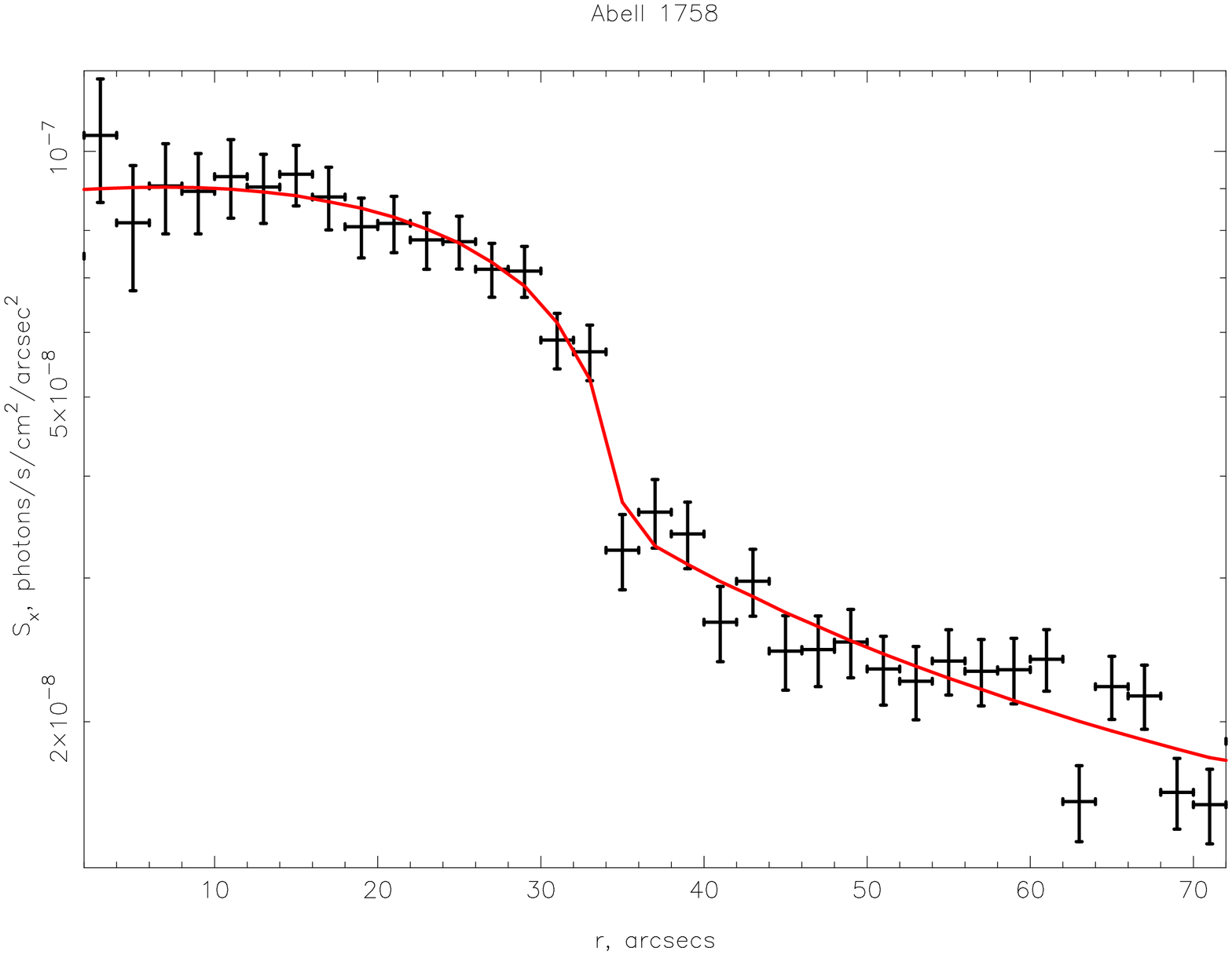}}
      {\includegraphics[angle=-90,width=0.31\textwidth]{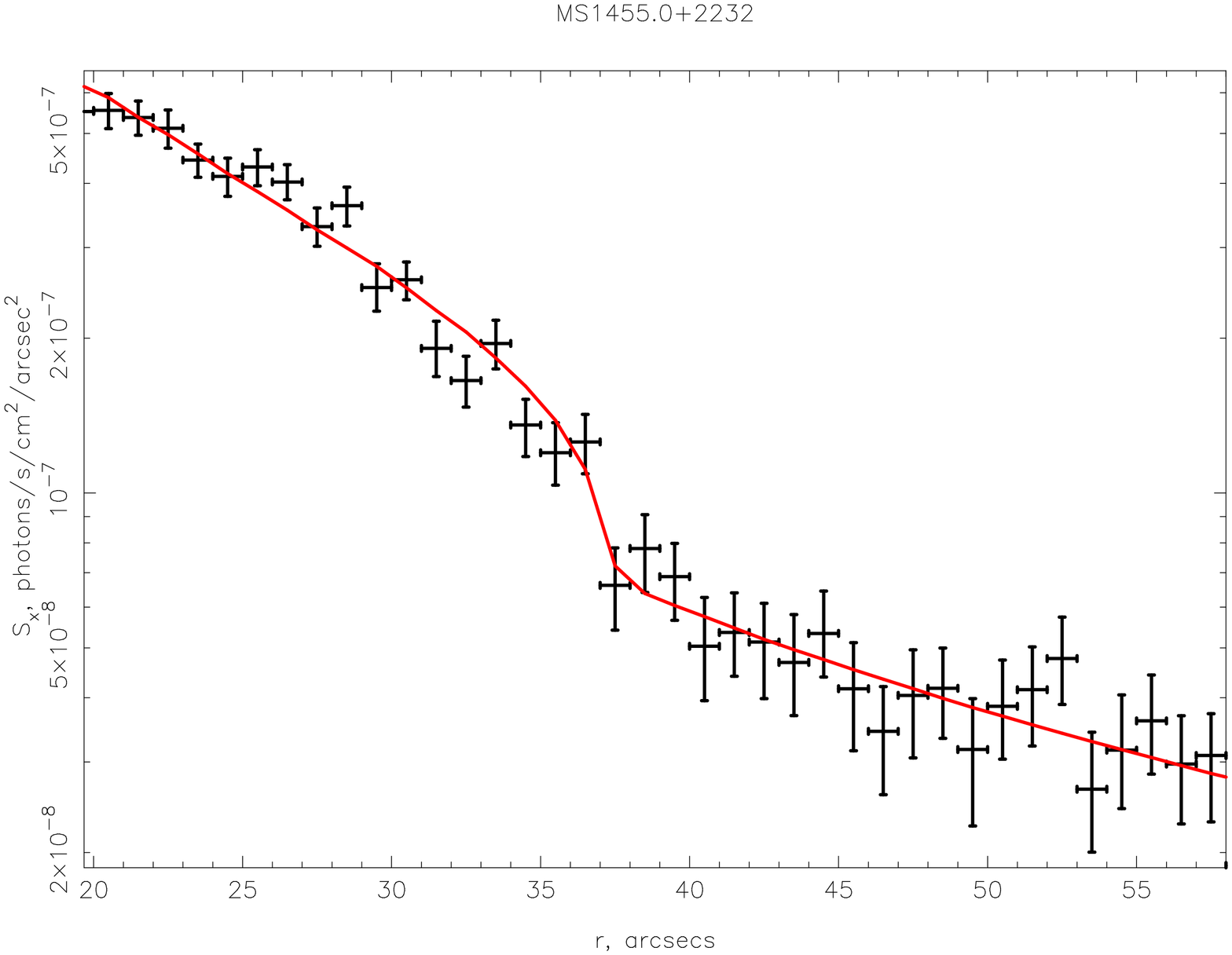}}
      {\includegraphics[angle=-90,width=0.31\textwidth]{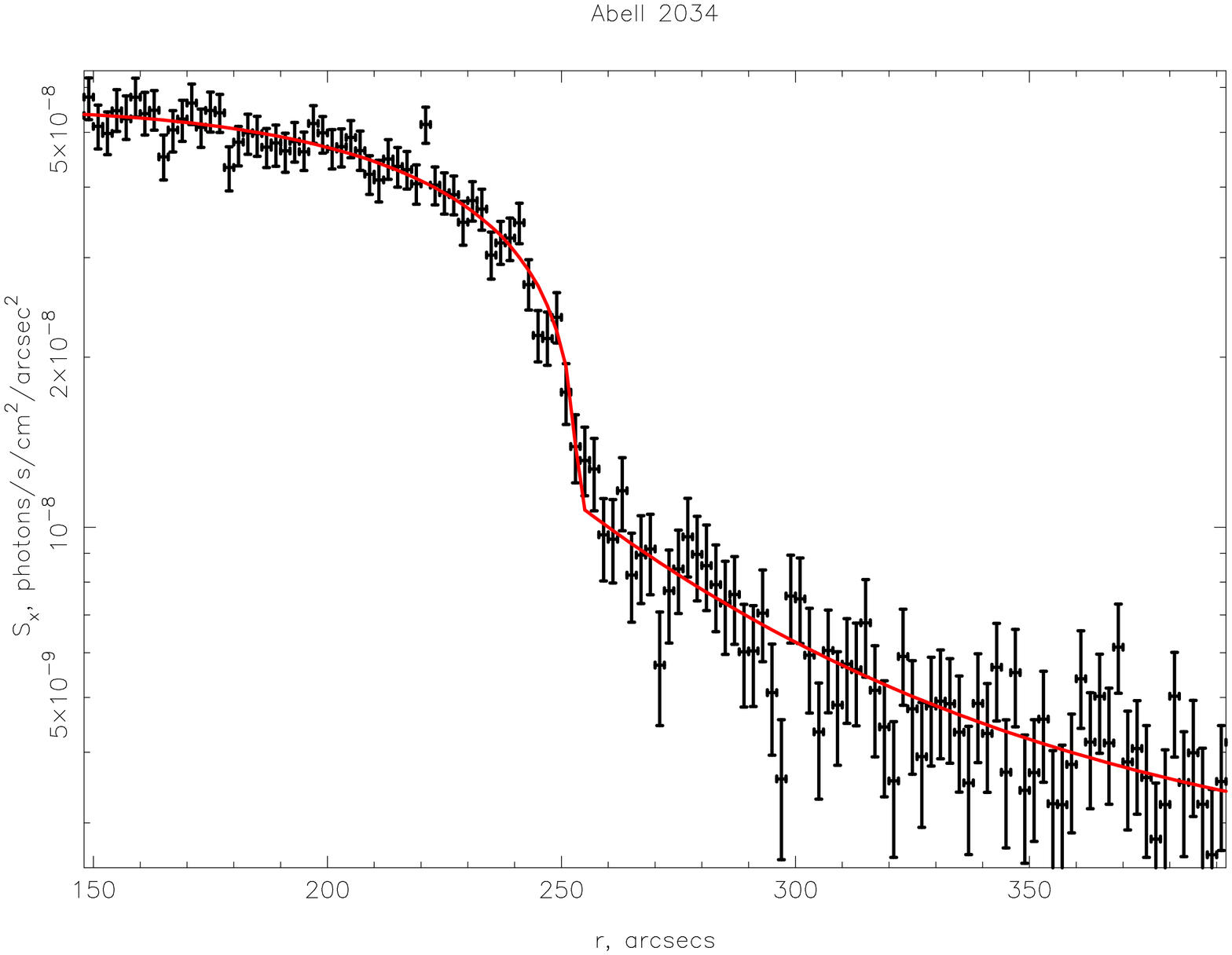}}\\
     {\includegraphics[angle=-90,width=0.31\textwidth]{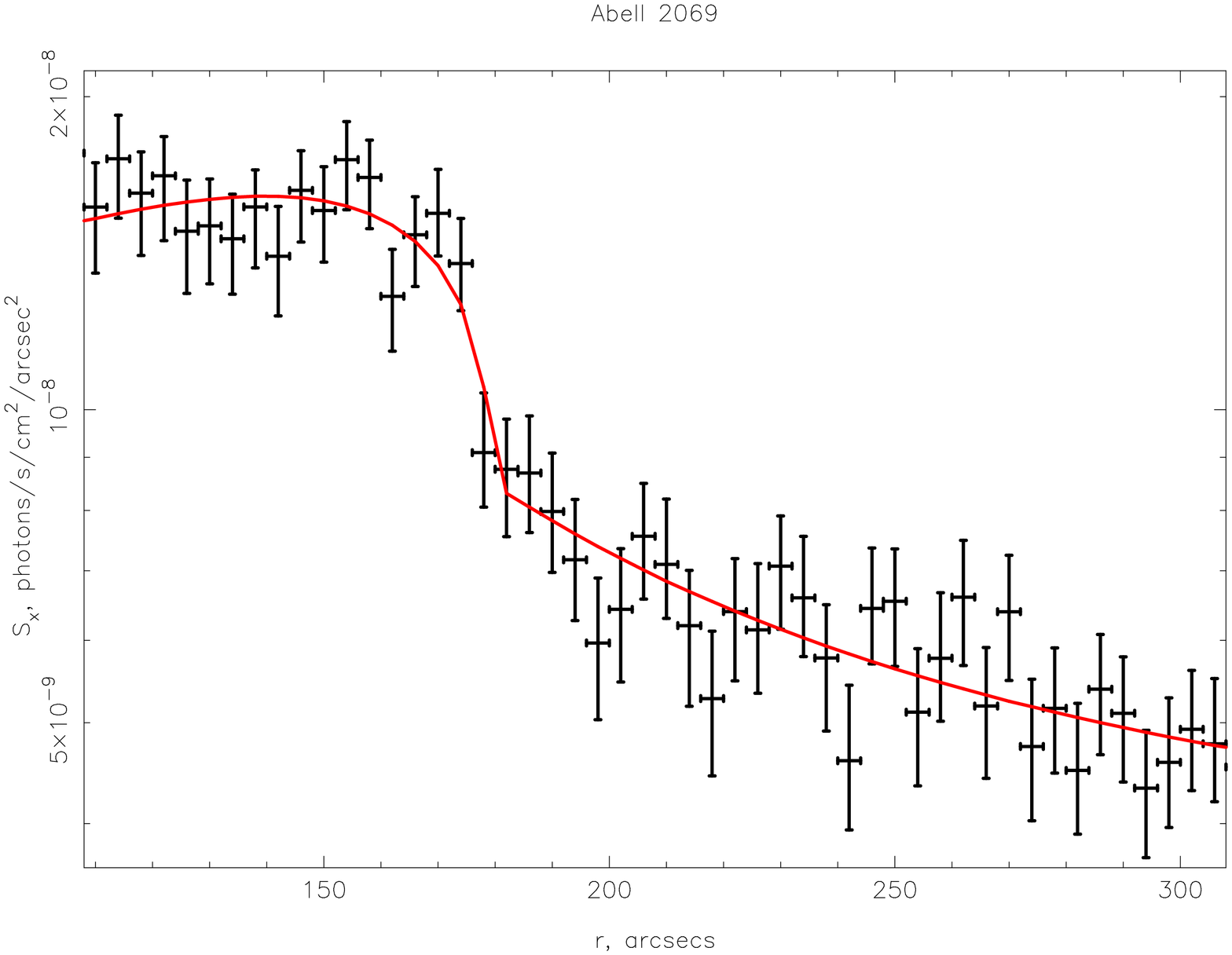}}
      {\includegraphics[angle=-90,width=0.31\textwidth]{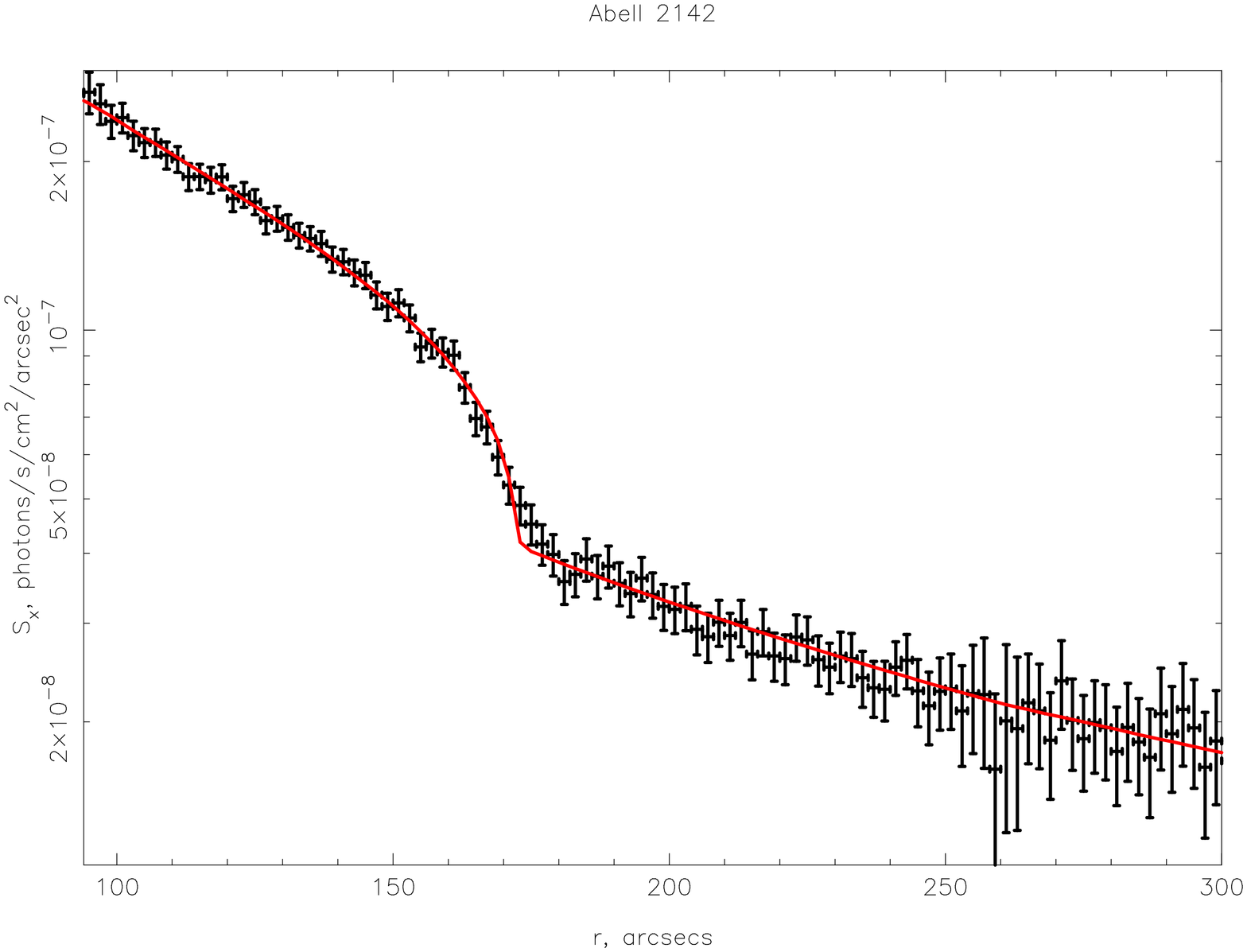}}
       {\includegraphics[angle=-90,width=0.31\textwidth]{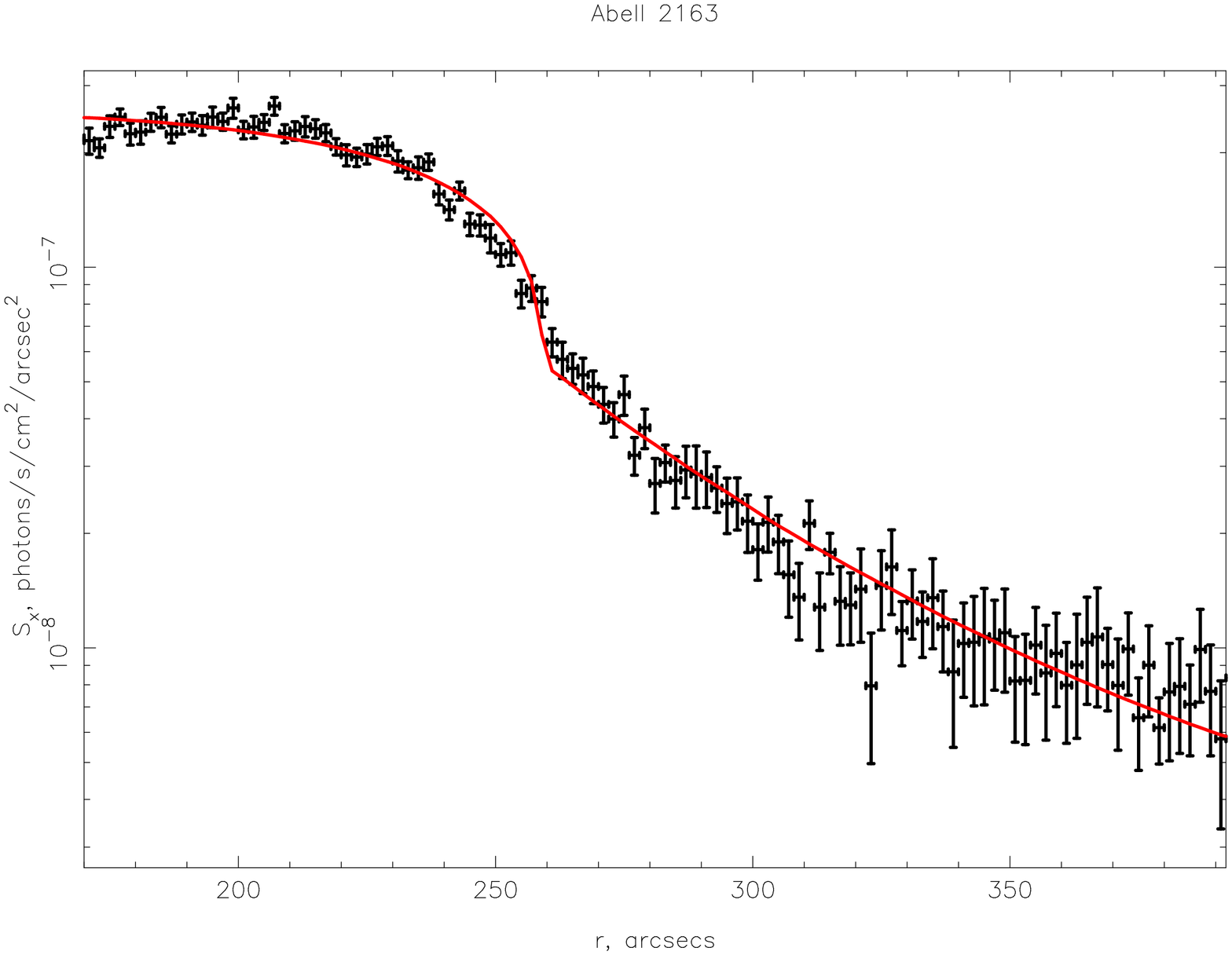}}\\
      {\includegraphics[angle=-90,width=0.31\textwidth]{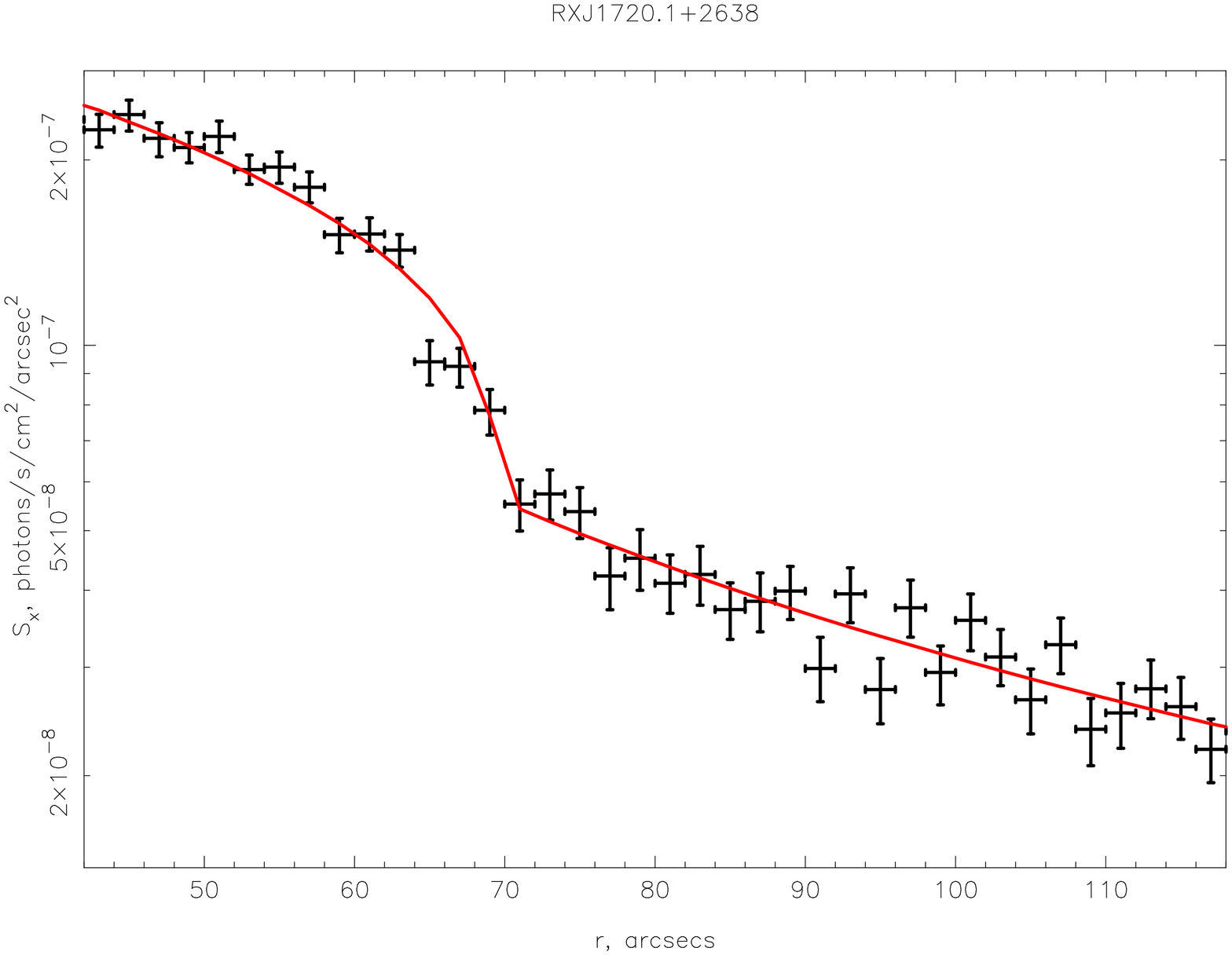}}
      {\includegraphics[angle=-90,width=0.31\textwidth]{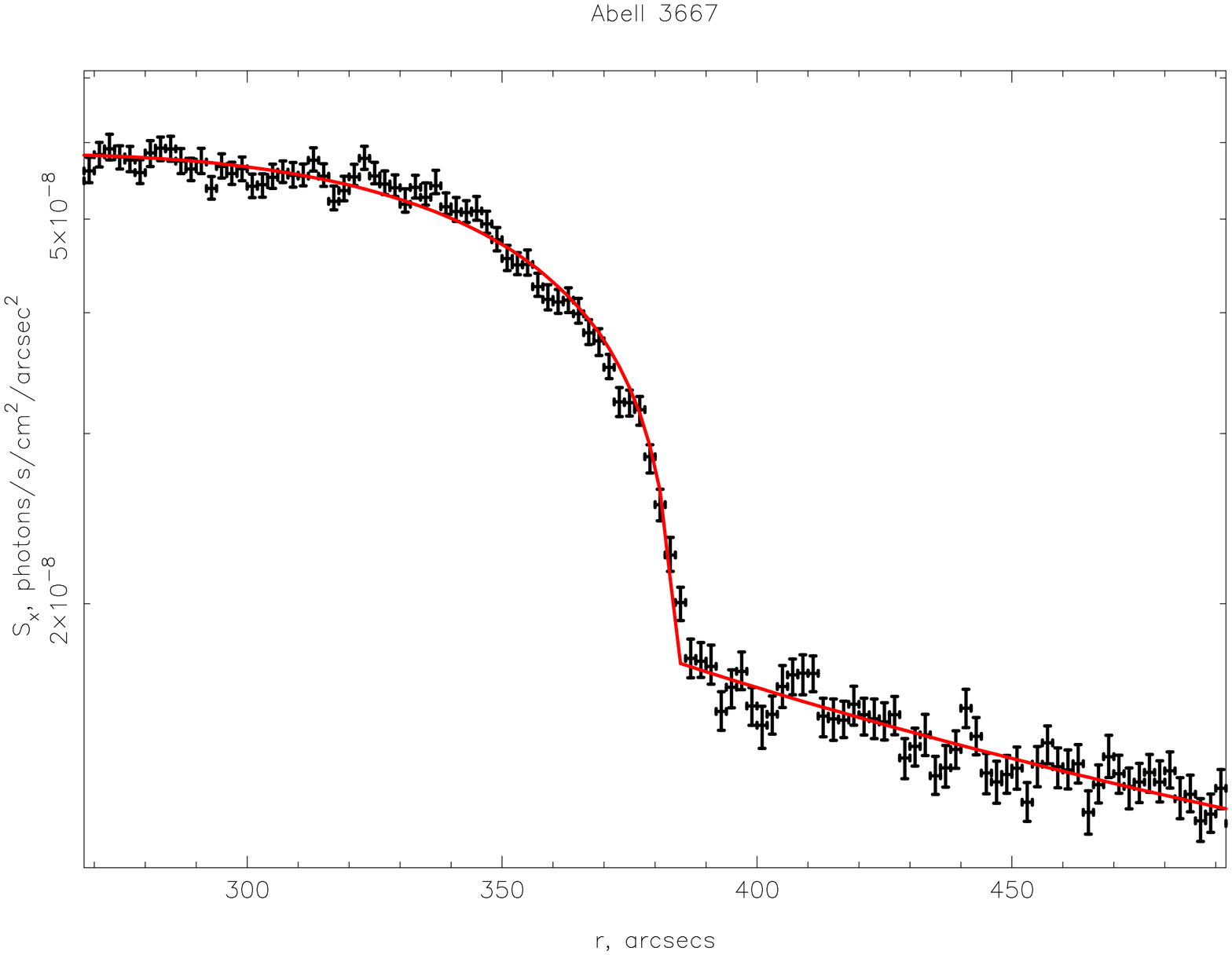}}      
    \caption{Surface brightness profiles across the edges (black 
points with error bars) with the surface brightness for the best fitting density
model plotted in red. The corresponding model parameters can be found in 
Table~\ref{sb.params}. The clusters shown are (from left to right, top row 
through bottom row) 1ES0657-558, Abell~665, Abell~1201, Abell~1758, MS1455.0+2232,
Abell~2034, Abell~2069, Abell~2142, Abell~2163, RXJ1720.1+2638 and Abell~3667.}
    \label{sb.profc}
  \end{center}
\end{figure*}

After fitting the density model (Appendix~\ref{sb.mod}), the sample was limited
to include only those clusters where the
density jump exceeded a factor of 1.5 at the lower limit of the $90\%$ confidence
interval. The confidence intervals were
estimated using the PROJECTION tool in {\it Sherpa}, which varies the parameter 
of interest along a grid of values whilst allowing adjustable model parameters 
to settle to their new best-fitted values. This criterion eliminated several 
clusters with clearly visible surface brightness edges from the sample.
Thus, any detectable cold fronts in clusters meeting the other selection 
criteria above that were missed in the search of the archive are unlikely to 
have density jumps exceeding a factor of 1.5. The 11 clusters which meet the 
above criteria, along with the parameters for the fits, are presented in 
Table~\ref{sb.params}. The corresponding surface brightness distributions with 
the surface brightness profiles for the best fitting density models overlaid are
plotted in Figure~\ref{sb.profc}. Interestingly, 
Abell~2142 is the only one of the 11 clusters to require a significant 
ellipticity in the fits. It is noted that density jumps
determined from the surface brightness fits do not depend on the
curvature of the front ($\epsilon_\zeta$ in Appendix~\ref{sb.mod}), but they do
depend on the assumption that the density profiles can be expressed in
terms of the same coordinate ($r$ in equation~\ref{sb_ne}) on both sides of the
front. The \chan\, Observation Identification number, cluster 
center and redshift are presented in Table~\ref{chap_cfsel:front.clusters} along
with global spectral properties (Section~\ref{global}). To confirm that these 
structures are cold fronts, temperature measurements on each side of the front 
are required (see Section~\ref{temp.jumps}).

\begin{deluxetable*}{cccccccccc}
\tabletypesize{\scriptsize}
\tablecolumns{10}
\tablewidth{0pt}
\tablecaption{Parameters of the density model fits for the 11 clusters with significant density discontinuities.\label{sb.params}}
\tablehead{
\colhead{Cluster} & \colhead{(x,y)} & \colhead{$\theta (\Delta \theta)$}  &\colhead{$R_f$} & \colhead{$\epsilon$} & \colhead{$A_1$} & \colhead{$\alpha_1$} & \colhead{$A_2$} & \colhead{$\alpha_2$} & \\
            &\colhead{(J2000)}&   \colhead{(deg)}    &\colhead{(kpc)} &            & \colhead{$(10^{-8})$\tablenotemark{a}} &&\colhead{$(10^{-8})$\tablenotemark{a}}&
}
\startdata
1ES0657-558 &104.58518, -55.941469 & 0  (310 - 50) & $53_{-0.2}^{+0.4}$ &  0 & $22.7_{-1.4}^{+1.5}$ & $0.0_{-0.1}^{+0.1}$ & $2.6_{-0.4}^{+0.4}$& $0.9_{-0.1}^{+0.1}$&\\
Abell~665 & 127.74663, 65.842256  & 252 (233 - 270) & $192_{-3}^{+5}$&  0 &$3.6_{-0.1}^{+0.1}$ &$0.7_{-0.4}^{+0.4}$ &$0.8_{-0.3}^{+0.3}$&$1.0_{-0.2}^{+0.2}$&\\
Abell~1201 & 168.2197, 13.4509 &254 (240-267) & $503_{-8}^{+4}$ & 0 & $2.0_{-0.3}^{+0.3}$ &$1.0_{-0.3}^{+0.4}$ &$0.4_{-0.1}^{+0.2}$ & $1.3^{+0.3}_{-0.2}$&\\
Abell~1758N & 203.1603, 50.5575 & 98 (45-150) & $150_{-2}^{+2}$&  0 &$3.9_{-0.4}^{+0.4}$ &$-0.1_{-0.1}^{+0.1}$ &$1.1_{-0.2}^{+0.2}$&$1.1_{-0.1}^{+0.1}$&\\
MS1455.0+2232 & 224.3125, 22.3427 & 132 (114-150) & $150_{-1}^{+1}$&  0 & $13.2_{-1.3}^{+1.4}$ &$1.2_{-0.2}^{+0.2}$ &$3.4_{-0.8}^{+0.9}$&$1.6_{-0.2}^{+0.3}$&\\
Abell~2034 & 227.53396, 33.495151  & 120 (104-126) & $521_{-2}^{+1}$&  0 &$3.7_{-0.2}^{+0.2}$ &$-0.4_{-0.1}^{+0.1}$ &$0.7_{-0.1}^{+0.1}$&$2.8_{-0.2}^{+0.2}$&\\
Abell~2069 & 231.12201, 29.994898 & 25  (3 - 46) & $383_{-3}^{+4}$ &  0 & $1.1_{-0.2}^{+0.2}$ & $-1.3_{->0.2}^{+0.5}$ & $0.3_{-0.1}^{+0.1}$& $1.8_{-0.2}^{+0.2}$&\\
Abell~2142 & 239.5863, 27.226989  & 40 (10 - 100) & $360_{-1}^{+3}$&  $0.36_{-0.01}^{+0.01}$ &$7.0_{-0.4}^{+0.1}$ &$0.9_{-0.03}^{+0.1}$ &$1.8_{-0.1}^{+0.1}$&$1.3_{-0.1}^{+0.04}$&\\
Abell~2163 & 243.97508, -6.1117719  & 315 (310-320) & $856_{-2}^{+2}$&  0 &$18.3_{-0.6}^{+0.7}$ &$-0.4_{-0.1}^{+0.1}$ &$5.2_{-0.4}^{+0.4}$&$3.8_{-0.1}^{+0.1}$&\\
RXJ1720.1+2638 & 260.03982, 26.624822  & 234 (205 - 263) & $190_{-1}^{+2}$&  0 &$10.1_{-0.9}^{+0.8}$ &$0.4_{-0.2}^{+0.2}$ &$2.4_{-0.4}^{+0.3}$&$1.4_{-0.1}^{+0.1}$&\\
Abell~3667 & 303.14776, -56.819424 & 230 (222 - 238) & $411_{-1}^{+0.3}$ &  0 & $4.4_{-0.1}^{+0.2}$ & $-0.7_{-0.1}^{+0.1}$ & $0.6_{-0.1}^{+0.1}$& $1.5_{-0.1}^{+0.2}$&\\
\enddata
\tablenotetext{a}{Units of $A_1\,{\rm and}\, A_2$ are $photons/cm^2/s/arcsec^2$}
\tablecomments{For a description of the density model parameters $A_1,\, A_2,\, \alpha_1$ and $\alpha_2$, see Appendix~\ref{sb.model}.}
\end{deluxetable*}

\subsection{Preparation of data for Spectral measurements}\label{spec.measurements}
Since the ACIS chips on board \chan\, record both spatial and energy information
for incident photons, it is possible to extract spatially resolved spectra from
the images. As for all astrophysical spectral observations, the data need to be
processed carefully in order to exclude events not related to the source (e.g. 
cosmic rays), calibrate the energy and also to account for the backgrounds. Here, 
the procedure used to prepare the data for spectral extraction is described. 

The archived \chan\ pipeline data were reprocessed starting with the level 1 
event files and using {\it CIAO} version 3.4 with the Calibration Database 
version 3.4.0. Observation specific bad pixel files were produced and applied, 
as were the latest gain files. Observations taken at focal plane
temperatures of $-120^\circ$C had corrections applied for charge transfer 
inefficiency and time dependent gain. Where observations were taken 
in VFAINT telemetry mode, the more thorough VFAINT mode of cleaning was used for
improved rejection of cosmic rays. Only events with {\it ASCA} grades of 0, 2, 
3, 4 and 6 were retained during the analysis. 

Backgrounds were taken from the blank sky observations appropriate for the 
epoch of observation\footnote{See http://cxc.harvard.edu/contrib/maxim/acisbg/}.
Since the blank sky backgrounds are taken from observations with low Galactic 
foregrounds and low soft X-ray brightness, the soft X-ray flux was checked in 
the vicinity of each cluster using the ROSAT all sky 
R45\footnote{See http://heasarc.gsfc.nasa.gov/docs/tools.html} count rates, and 
compared to the blank sky background rates. If the R45 fluxes were anomalous
(e.g. MS1455.0+2232, Abell~2163 and Abell~3667),  a region relatively free from
cluster emission was selected and spectra extracted to model the soft component
using a single or double MEKAL model at zero redshift 
\citep[see e.g.][]{markevitch2003}. The soft component was included in all 
spectral analyses, with the normalization weighted by a geometric factor 
to allow for the different extraction areas. The background and source data were 
processed using the same calibration files, bad pixel files and background 
filtering, with the backgrounds being reprojected to match the observations. 

To match the blank sky backgrounds, the observations were filtered for periods 
of high cosmic ray background caused by flares. The point sources and regions 
containing cluster emission were excluded and a light curve was extracted and 
binned in time intervals of 259 seconds in the energy range 0.3-10 keV for the 
ACIS-I array, whilst the ACIS-S array data were binned in intervals of 1037 
seconds in the energy range 2.5-6 keV. The light curve was analyzed in {\it 
Sherpa} using the {\sf lc\_clean.sl} script and only periods where the 
background count rate was within 20\% of the quiescent rate were included.

During the spectral analysis of Abell~1758N, it was necessary to account for a 
soft flare present for the duration of the observation. The method described in
\citet{david2004} was used, whereby the flare shape was modeled in PHA space 
in a region relatively free from cluster emission, using the {\sf cutoffpl} model
in XSPEC with power law index = 0.15 and an exponential cut off at E=5.6 keV, 
with the normalization allowed to vary. The flare model was then incorporated 
into spectral fits with all parameters fixed and the normalization set to its 
best fitted value from above multiplied by a geometric factor correcting for the 
different extraction region areas.

\subsubsection{Global Temperatures}\label{global}
To obtain a picture of the overall thermal properties of the 11 clusters, 
global temperatures and metallicities were measured. Where possible, a region 
covering the majority of the cluster emission was selected and spectra were 
extracted using the {\it CIAO} DMEXTRACT tool, with point sources detected with 
WAVDETECT excised. The regions within which the spectra were extracted are shown
in red in the X-ray images presented in the top left panels of 
Figures~\ref{fig:1e657a} through \ref{fig:a2142a}. The {\it CIAO} tool MKWARF 
was used to create auxiliary response files (ARF), which account for spatial 
variations in quantum efficiency (QE), effective area and also the temporal 
dependence of the effective area due to molecular contamination on the optical 
blocking filter. For observations performed with a focal plane temperature of 
$-120^\circ$C, the {\it CIAO} MKACISRMF tool was used to create redistribution 
matrix files (RMF), while MKRMF was used for this purpose for observations 
performed at $-110^\circ$C. Since the responses depend on the chip position, 
they are calculated in regions in chip coordinates and are weighted by the 
0.5 -- 2 keV brightness of the cluster within that region, with the final ARF 
and RMFs being count weighted. For clusters with multiple observations where a 
more recent, significantly longer exposure was available, the earlier, shorter 
observations were excluded from the global temperature measurement.

The spectra were fitted in the 0.5-9.8 keV range for ACIS-S observations and 
0.6-9.8 keV for ACIS-I observations in XSPEC \citep{1996ASPC..101...17A} using 
a MEKAL model \citep{mewe1985, mewe1986, kaastra1992,1995ApJ...438L.115L}, for a hot, diffuse, 
single temperature plasma, multiplied by WABS, a photo-electric absorption 
model accounting for Galactic absorption, where the neutral hydrogen column 
density is initially fixed to the Galactic value provided by 
\citet{1990ARA&A..28..215D}. The spectra were binned so that each energy bin 
contains at least 30 counts and the normalization, temperature and abundance 
were fitted in the first iteration, with the column density fixed to the Galactic
value. The metal abundances are measured relative to the solar photospheric 
values of \citet{anders1989}. Where the data allow, the column density was then 
freed and the data re-fitted, with the best fitting results along with their 
associated $90\%$ errors presented in Table~\ref{chap_cfsel:front.clusters}. 
For the clusters 1ES0657-558, Abell~2142, Abell~2163 and Abell~3667 the measured 
$N_{\rm H}$ values are inconsistent with the Galactic values within the $90\%$ 
confidence limits, although they are consistent with values measured by others 
in the literature \citep[e.g.][]{govoni2004,markevitch2000,vikhlinin2001b}. In 
the clusters Abell~665 and \rxj, the $N_{\rm H}$ values are also inconsistent with 
Galactic values, however the temperature and abundance  measurements appear to 
be relatively unaffected and they are consistent with previously measured values
\citep[][for \rxj\, and Abell~665, respectively.]{mazzotta2001,govoni2004}. The
LUMIN function in XSPEC was used to determine the integrated, unabsorbed X-ray 
luminosity from the best fitting model in the energy range 0.5-9.8 keV, with the
results presented in Table~\ref{chap_cfsel:front.clusters}.

\begin{deluxetable*}{lccccccccc}
\tabletypesize{\scriptsize}
\tablewidth{0pt}
\tablecaption{Clusters with density jumps exceeding 1.5 at the lower $90\%$ 
confidence interval. The exposure times presented in parentheses are the 
cleaned exposure times used to generate the temperature maps. The $N_{\rm H}$ 
values presented in parentheses are the galactic values.\label{chap_cfsel:front.clusters}}
\tablehead{
\colhead{Cluster} & \colhead{ObsId} & \colhead{Exposure} &\colhead{RA} & \colhead{DEC}  & \colhead{Redshift} & \colhead{$kT$}& \colhead{$Z$ ($Z\odot$)} & \colhead{$L_{\rm X}$}  &\colhead{$N_{\rm H}$} \\
&&\colhead{(ks)}&\colhead{(J2000)}&\colhead{(J2000)}&&\colhead{(keV)}&&\colhead{($10^{44}\rm\ erg/s$)}&\colhead{$\times10^{20}$}
}
\startdata
1ES0657-558&3184, 4984, 4985,&551 (535) &06:58:27&-55:56:47&0.296&$14.7^{+0.3}_{-0.3}$&$0.32^{+0.03}_{-0.02}$&43.0&$3.7^{+0.3}_{-0.3}$(6.53)\\
&4986, 5355, 5356&&&&&\\
&5357, 5358, 5361&&&&&\\

Abell 665 & 531, 3586 &40 (35)&08:30:45 & 65:52:55 & 0.182 & $8.5^{+0.4}_{-0.7}$ &$0.30^{+0.10}_{-0.05}$&13.6& $2.7^{+1.2}_{-0.8}$(4.24)\\

Abell 1201 & 4216 &40 (22) &11:13:01 & 13:25:42& 0.168 & $5.3^{+0.3}_{-0.3}$&$0.34^{+0.10}_{-0.10}$&4.00&1.61 \\

Abell 1758N & 2213 &58 (42) &13:32:32 & 50:30:36 & 0.279 & $8.4^{+0.5}_{-0.5}$&$0.56^{+0.12}_{-0.11}$ &14.2 &1.05\\

MS1455.0+2232 & 534, 4192 &102 (90)  &14:57:15 & 22:20:30 & 0.258 &$4.5^{+0.1}_{-0.1}$&$0.43^{+0.04}_{-0.04}$& 14.6&3.1\\

Abell 2034& 2204&54 (53) &15:10:13 & 33:31:42 & 0.113 & $6.4^{+0.2}_{-0.2}$ &$0.26^{+0.05}_{-0.04}$&5.4&1.58 \\

Abell 2069 &4965& 55 (39) &15:23:58 & 29:53:26 & 0.116 & $6.2^{+0.3}_{-0.3}$ &$0.28^{+0.07}_{-0.07}$&4.6&1.96 \\

Abell 2142 &1196, 1228, 5005&68 (45) &15:58:16 & 27:13:29 & 0.089 & $9.8^{+0.3}_{-0.2}$ &$0.41^{+0.02}_{-0.02}$&17.5&$3.3^{+0.3}_{-0.4}$(4.20) \\

Abell 2163 &545, 1653&81 (80) &16:15:34 & -06:07:26 & 0.201 & $15.3^{+0.8}_{-0.8}$ &$0.26^{+0.04}_{-0.05}$&34.1&$15.4^{+0.8}_{-0.4}$ (12.1)\\

RXJ1720.1+2638 & 304, 549, 1453& 60 (50)&17:20:09 & 26:37:35 & 0.164 & $6.1^{+0.3}_{-0.1}$&$0.41^{+0.05}_{-0.03}$&14.0 &$3.2^{+0.6}_{-0.9}$(4.06)\\
&3224, 4361&&&&&\\
Abell 3667 & 513, 889, 5751& 534 (473) &20:12:34&-56:50:26&0.055&$7.1^{+0.1}_{-0.1}$&$0.40^{+0.01}_{-0.01}$&3.9&$4.22^{+0.2}_{-0.2}$ (4.71)\\

&5752, 5753, 6292&&&&&\\
&6295, 6296&&&&&\\
\enddata
\end{deluxetable*}

\subsubsection{Temperatures across the edges}\label{temp.jumps}
Cold fronts are expected to have very different thermal properties when 
compared to shock fronts. The pressure ($P\sim n_e kT$) across a cold front 
should be continuous, whilst a shock front should show a pressure discontinuity.
The specific entropy, specified here in terms of the entropy index, 
$\Sigma= kTn_e^{-2/3}$, should also change rapidly across a cold front
since, regardless of the physical mechanism causing the cold front, the front 
occurs at the interface of cool, dense low entropy gas and the hotter, ambient 
higher entropy ICM. Thus, a jump in the entropy across a front is expected which
will also be opposite to the entropy change expected across a shock front. To 
ensure the observed surface brightness edges are indeed cold fronts, 
the temperature was measured just inside and outside the edges and used
in combination with the density jumps measured above to derive changes in 
pressure and entropy across the edges. 

The precision of the pressure measurements across the fronts is determined by
limitations in both the data and the assumptions used in modeling the surface 
brightness across the edges to obtain densities. For example, in obtaining 
densities it has been assumed that the gas density is a function of the 
elliptical radius. This is an approximation, which may be poor if, e.g., there 
is emission from gas projected onto the edge region, or there are rapid, 
nonradial variations in the gas distribution. These effects are reduced by 
confining attention to the most prominent fronts, where the X-ray emission is 
generally dominated by gas close to the front. Furthermore, while pressure 
continuity holds in the immediate vicinity of the cold front, if there is a 
stagnation region at the front then in upstream regions away from the cold 
front, where the velocity gradient is high, the pressure will vary 
\citep[see Figure 6 of][]{vikhlinin2001b}. Thus, pressure measurements must be 
made in regions as close to the front as possible. The size of these regions is
most affected by the temperature measurements, which require a large number of 
counts in order to be accurate, meaning the precision of the pressure 
measurement across the edges is limited mostly by the temperature measurements, 
rather than the shortcomings of the surface brightness model.

In order to minimize the above effects, regions from which spectra were 
extracted for temperature measurements are chosen to lie as close as possible 
to the fronts. The opening angles for these regions correspond to those 
used for measuring the surface brightness profiles in Section~\ref{sb.fit} (see 
the blue regions plotted in the left panel of Figures~\ref{fig:1e657a} through 
\ref{fig:a2142a}). Three exceptions exist in Abell~1201, Abell~665 and 
MS1455.0+2232 where for the region just outside the edge, a slightly larger 
region with a larger angular extent was used (MS1455.0+2232 has a slightly 
larger radius compared to the region where the surface brightness profile 
was measured, too), so that there are enough counts to measure the temperature 
with sufficient accuracy. Using a larger angular extent was preferred to 
increasing the radial extent for the reasons discussed above. Regions were 
chosen with a minimum of 700 counts in the 0.5 -- 7\,keV energy range after 
background subtraction. The green arcs shown in top left panels of 
Figures~\ref{fig:1e657a} through \ref{fig:a2142a} show the inner and outer 
limits for the regions used to measure the temperatures on the inside and 
outside of the edges, respectively. Where a green arc is not plotted, the 
radial limit used corresponds to that used in measuring the surface brightness 
profile (i.e. the blue regions).

The procedure for fitting the temperature inside and outside the edge 
was as follows. First, the temperature in the region outside the edge 
was fitted using an absorbed MEKAL model with the column density and 
metallicity fixed\footnote{We have tested the 
effect of allowing the metallicity to vary during the fitting procedure using the
Abell~3667 observations. We find that, within the errors, the temperature 
measurements are not affected, despite measuring metallicities of 
${Z_{1}}=0.84^{+0.27}_{-0.22}$ and ${Z_{2}}=0.49^{+0.33}_{-0.27}$ inside and outside 
the front, respectively.} to the values derived in Section~\ref{global}. Second, 
the spectrum from the region just inside the edge was fitted with two absorbed MEKAL
models. The first MEKAL component accounts for the denser parcel of gas lying 
within the edge. The second MEKAL component accounts for gas lying in projection
along the line of sight which is assumed to have the same thermal properties as 
the gas fitted in the region outside the edge. Thus, the temperature of
this second MEKAL component was fixed to that measured for the region outside 
the edge. In XSPEC, the normalization of the MEKAL model is defined as 
\begin{equation}\label{norm}
Norm = {10^{-14}\over {4 \pi D_A(1+z)^2}}\int n_e n_p dV , 
\end{equation}
where $D_A$ is the 
angular diameter distance to the cluster, $n_e$ and $n_p$ are the electron and 
proton densities. This is proportional to the emission measure, so the 
normalization of the second MEKAL component was fixed to the value derived for 
the outer region, however it was corrected by a factor accounting for the 
different emission measures expected from the different volumes probed.  Using 
the density model for $n_e(r)$ in Equation~\ref{sb_ne} for $R_F<r<R_{out}$ 
(where $R_{out}$ is the outer radius of the region used to measure the 
temperature outside the front, shown in each cluster's
\chan\ image [left panel of Figures~\ref{fig:1e657a} through \ref{fig:a2142a}]),
along with the parameters presented in Table~\ref{sb.params}, 
Equation~\ref{norm} can be integrated for both regions inside and outside the 
edge and the ratio used to obtain the correction factor applied to the 
normalization of the second MEKAL component.

\begin{deluxetable*}{ccccccc}
\tabletypesize{\scriptsize}
\tablewidth{0pt}
\tablecaption{Properties across the edges in surface brightness.\label{front.props}}
\tablehead{
\colhead{Cluster} & \colhead{Density Jump} & \colhead{$kT$ Inside} & \colhead{$kT$ Outside} & \colhead{Pressure Jump} & \colhead{Entropy Jump} & \colhead{$kT_2$ for shock}\\
       &\colhead{($\sqrt{A_1\over A_2}$)}& \colhead{($kT_1$)}     &  \colhead{($kT_2$)}      & \colhead{$({n_{e,1}\over n_{e,2}})({kT_1\over kT_2})$}& 
\colhead{$({kT_1\over kT_2})({n_{e,2}\over n_{e,1}})^{2/3}$}  &
}
\startdata
1ES0657-558& 3.0(2.7-3.4) & 5.8  (5.3-6.3)& 19.7(16.9-22.2) & 0.9 (0.6-1.3)& 0.1 (0.1-0.2)&1.6 (0.9-2.4)\\
Abell~665& 2.2(1.6-3.2) & 9.2 (7.1-13.1) & 8.6 (6.6-11.9) & 2.3 (1.0-6.3) & 0.6 (0.3-1.4)&4.8 (1.6-9.1)\\
Abell~1201 & 2.1(1.7-2.7) & 3.6 (2.9-4.7) & 5.7 (4.0-9.4)&1.4 (0.5-3.2) & 0.4(0.2-0.8)&1.9 (1.0-3.2 )\\
Abell~1758N & 1.9(1.7-2.2) & 7.9 (5.7-12.1) & 10.3 (7.8-15.4)& 1.5 (0.6-3.4) & 0.5 (0.2-1.1)&4.8 (2.9-8.3 )\\
MS1455.0+2232  & 2.0(1.7-2.4) & 3.7 (3.3-4.3) & 7.0 (5.6-9.2)& 1.0 (0.6-1.8)& 0.3 (0.2-0.6)&2.1 (1.5-3.0 )\\
Abell~2034 & 2.3(2.1-2.5) & 6.6 (4.9-10.2) & 4.8 (3.9-6.3) & 3.1 (1.6-6.6) & 0.8(0.4-1.6)&3.2 (2.0-5.6)\\
Abell~2069 & 2.0(1.7-2.4)& 4.5 (3.5-6.3) & 6.0 (4.4-9.1)& 1.5 (0.6-3.4) & 0.5 (0.2-1.0)&2.6 (1.6-4.4)\\ 
Abell~2142 & 2.0(1.9-2.1) & 7.5 (6.1-9.5)& 17.6 (13.1-25.4) & 0.8 (0.5-1.5)& 0.3 (0.2-0.5)&4.3 (3.3-5.7)\\
Abell~2163 & 1.9(1.8-2.0) & 8.6 (7.1-10.8) & 13.2 (9.9-18.9) & 1.2 (0.7-2.2)& 0.4 (0.24-0.8)&5.2 (4.1-7.0)  \\
RXJ1720.1+2638 & 2.1(1.9-2.3) & 5.4 (4.4-6.8)& 7.4 (6.3-9.3)& 1.5(0.9-2.5) & 0.5 (0.3-0.7)&1.8 (1.6-2.1)\\
Abell~3667 & 2.6(2.4-2.8) & 3.7 (3.5-4.0) & 8.2 (7.3-9.3) & 1.2 (0.9-1.6)& 0.2 (0.2-0.3)&1.4 (1.2-1.8)\\
\enddata
\end{deluxetable*}

Table~\ref{front.props} lists the results of the temperature measurements 
across the edges, as well as the density, entropy and pressure jumps for 
each cluster in the sample. For comparison, the temperature expected on the less
dense side of the edge is computed under the assumption that the density
jump is caused by a shock front using the Rankine-Hugoniot condition
\begin{equation}
{kT_2 \over kT_1} = {{(\gamma+1)-{n_{e,1} \over n_{e,2}}(\gamma-1)}
\over{(\gamma+1)-{n_{e,2} \over n_{e,1}}(\gamma-1)} }
\end{equation}
where $\gamma=5/3$ is the adiabatic index for a monatomic gas, quantities with
subscript 1 refer to those on the inside of the edge (i.e. denser side),
and those with subscript 2 refer to quantities on the outside of the edge (less 
dense side). Abell~2034 and Abell~665 both have best fitting $kT_2$ values which
are lower than $kT_1$ and, although the error bars would allow the temperature 
to be continuous across the front, the defining criterion of pressure continuity
across a cold front is violated in these clusters (although for Abell~665 the 
lower error bar marginally allows for pressure continuity). A more likely 
interpretation for these edges in surface brightness is that of shock 
fronts, and this can be also seen by comparing the expected temperatures for the
pre-shock gas given by the Rankine-Hugoniot shock conditions with the 
temperature measured on the outside of the edge. Thus, Abell~2034 and 
Abell~665 are excluded from the cold front sample. However, shock fronts 
strongly imply the existence of merger activity within these clusters and 
contamination in a cold front sample by shock fronts does not have a 
detrimental effect if the primary objective is to use the fronts as signposts 
for merger activity. 

The remaining nine clusters all have thermal properties across their surface 
brightness edges which are consistent with that expected for a cold 
front. The cold front sample is comprised of these nine clusters and their X-ray
images and projected temperature maps are presented in Section~\ref{cf.clusters}.

\section{Temperature maps}\label{tempmaps}
Temperature maps can offer further insights into the overall dynamical state of
a cluster, revealing regions which have been heated by shocks, dissipation of 
turbulence and adiabatic compression during a merger, and are also useful for 
comparing to the temperature structures found in simulations of cluster mergers.
For the cold front clusters, temperature maps were created using the broad 
energy band method described in \citet{markevitch2000}. Briefly, for each 
cluster, source images were produced in several energy bands in the range 
0.5 -- 10\,keV. The energy bands were chosen to include at least 5000  
counts per band (or 10000 where the data allow). Exposure maps correcting for 
mirror vignetting, QE variations, photoelectric absorption by contaminant on the
filters and exposure time were produced for each energy band. Background images 
in the corresponding energy bands were produced using the blank sky observations
described in Section~\ref{spec.measurements} and were normalized by the ratio of
the source to background 9-12 keV counts and subtracted from the source images 
which were then divided by the exposure maps. Each image was smoothed using the 
same variable width Gaussian, where the width of the Gaussian, $\sigma(r)$, is 
smallest in the brightest regions, i.e. the regions with the most source counts,
and becomes larger as the brightness lowers. Different initial values of 
$\sigma(r)$ were trialled until a value was found which allows statistically 
significant temperatures to be measured (i.e., the 1$\sigma$ errors did not 
exceed $30\%$ of the best fitting value at each pixel) whilst maintaining a good
degree of spatial resolution. The noise in each pixel was determined from the 
raw, uncorrected images and weighted accordingly to allow for the effects of the
smoothing. 

The temperature maps were produced by fitting an absorbed MEKAL model to the
band fluxes for each pixel \citep{markevitch2000}. The absorption column was 
set to either the Galactic value, or the best fitting value from 
Section~\ref{global}. The metal abundance was set to the best fitting average 
cluster value derived in Section~\ref{global}. The model was multiplied by an 
on-axis ARF to correct for the energy dependent on-axis mirror effective area, 
including the chip quantum efficiency (QE). A flux-weighted spectral response 
matrix was generated from a large cluster region, and was binned to match the 
chosen energy bands. 

Where there were multiple observations for a single cluster, images and exposure
maps were co-added to produce single images in each energy band. However, if 
there was a significant time gap between observations, or the observations were 
taken on different chips, separate images were produced for each observation in 
each energy band and the data were simultaneously fitted to produce a single 
temperature map. Regions in the temperature map where the $1\sigma$ errors were 
greater than $30\%$ of the best fitting temperature, or where the best fitting 
temperature exceeds 25 keV, were excluded from the maps which are shown in 
the top right panel of Figures~\ref{fig:1e657a} through \ref{fig:a2142a}. Regions near
the edges of the maps, where the counts are low, are also excluded. Furthermore, 
for Abell~1758N, the spatial variations of the soft flare
are not accounted for in the generation of the temperature map. However, the 
temperature structure seen in the right panel of Figure~\ref{fig:a1758a} agrees 
qualitatively with structure seen in maps which have been corrected for this 
spatial nonuniformity\footnote{See http://cxc.harvard.edu/ccr/proceedings/03\_proc/prese-
ntations/markevitch2/index.html}. As we are interested in qualitative 
temperature structure, we chose not to include modeling of the flare spatial 
structure here for simplicity and note that our conclusions are not affected.

\section{Notes on individual clusters}\label{cf.clusters}
Here, the X-ray images, temperature maps and $r$-band optical images for each of the cold 
front clusters are presented and discussed. There is a clear dichotomy in the
sample when considering X-ray morphology insomuch that the
sample can be divided into those clusters which appear disturbed (i.e. they are 
clear mergers with a complex appearance and/or substructure) and those which, 
aside from the existence of a cold front, appear relaxed. To highlight this, the
sample is partitioned based on X-ray morphology with Section~\ref{disturbed} 
containing the disturbed clusters and Section~\ref{relaxed} containing the 
relaxed appearing clusters. Within these sections, the clusters are arranged in 
decreasing order of density jump strength. For each cluster, we discuss 
further evidence for merger activity present in the literature and, where 
possible, relate this to the existence of the cold front.

It is noted that all of the clusters in the sample have previously published 
work based on \chan\ data (aside from Abell~2069), and some of these studies 
included temperature maps of the cold front clusters. However, for the purpose 
of comparison, it is best to present temperature maps which have been produced 
in a homogeneous manner. In addition to this, many of the clusters have since 
been reobserved by \chan\, (for example, Abell~2142 and Abell~3667) meaning 
the maps produced here are of a higher statistical quality and resolution than 
those previously published. Where pertinent, the temperature maps presented here
are compared to the previously published versions. For each temperature map, the
color scale used was chosen so that different colored regions reflect regions 
where temperatures differ by at least $\sim 1\sigma$.

\subsection{Clusters with Disturbed X-ray Morphologies}\label{disturbed}

\subsubsection{1ES0657-558 - Bullet cluster}
\begin{figure*}
  \begin{center}
     {\includegraphics[width=\textwidth]{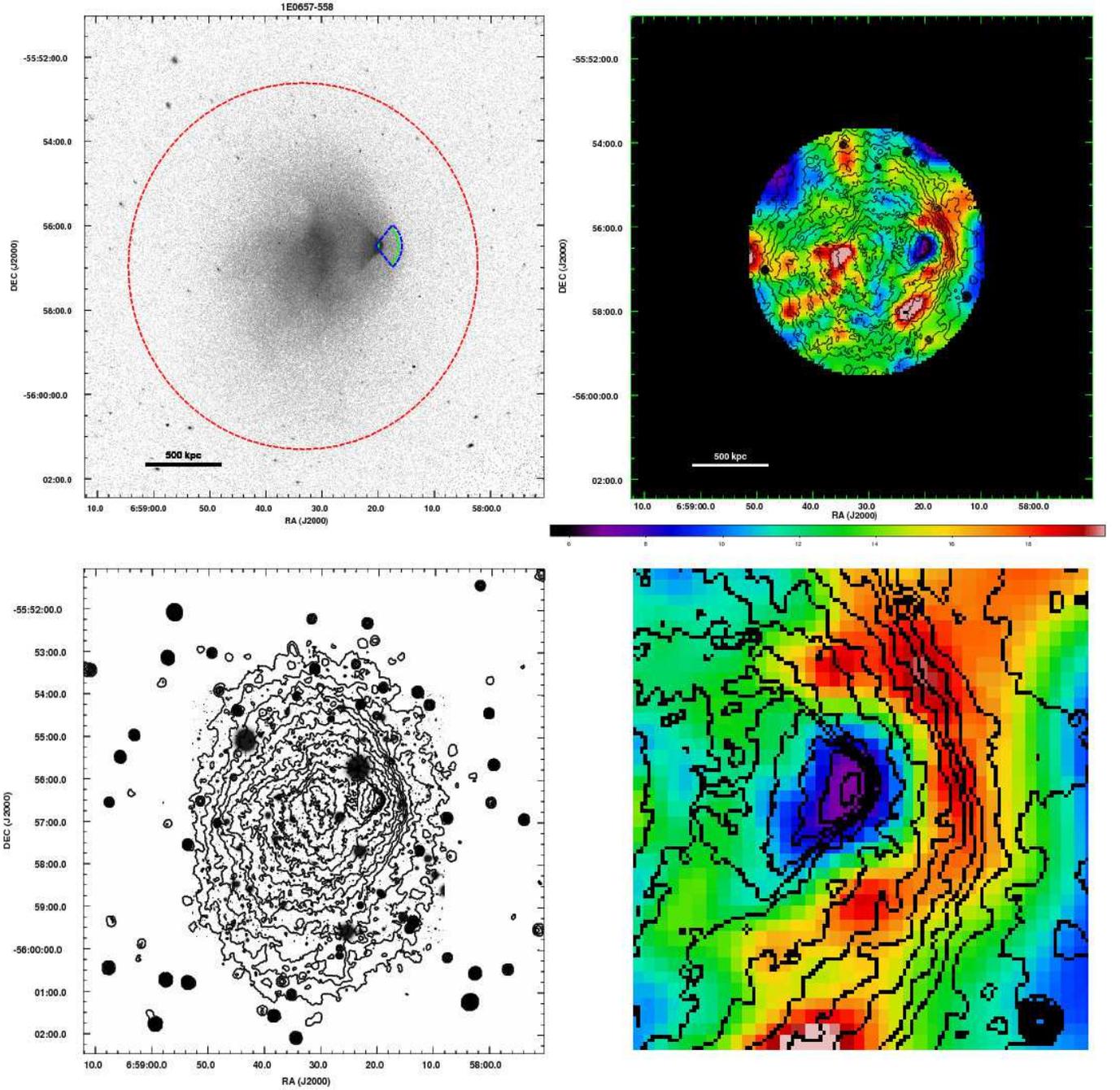}}
\caption{Top left: background subtracted, exposure corrected, 0.5 -- 7\,keV \chan\, 
image of  1ES0657-558. The red region shows the area used to measure the global temperature 
in Section~\ref{global}, and the blue region is where the surface brightness 
profile was fitted. The green arcs show the radial limits used for measuring 
the temperature inside and outside the cold front (the cold front delineates 
the two regions). North is towards the top of the page, and East is to the left.
Top right: this image shows the projected temperature map produced as described
in Section~\ref{tempmaps}. The color scale is tuned so that different colors 
correspond to regions where temperatures differ by approximately $1\sigma$. 
Bottom left: Magellan IMACS $r$-band image (courtesy of D. Clowe). Bottom right:
zoomed in version of the temperature map showing the region containing the cold
front. The top right, bottom left and bottom right images have X-ray surface 
brightness contours overlaid.}  
      \label{fig:1e657a}
  \end{center}
\end{figure*}

1ES0657-558 was discovered serendipitously as an extended X-ray source in the 
{\it Einstein} Slew Survey \citep{elvis1992} and was later found to be one of 
the hottest, most luminous clusters known \citep{tucker1998} with redshift 
$z=0.296$ and a velocity dispersion of $\sigma_v \simeq 1200$\kms 
\citep{barrena2002}. \citet{markevitch2002} presented the first \chan\ 
observations of 1ES0657-558, which revealed a high average temperature of 
$\sim 15$\,keV and a clearly disturbed morphology 
consisting of a main cluster, along with a subcluster to the west 
and two sharp surface brightness edges on the western side of the 
subcluster. These features are seen in greater detail in the \chan\ image
presented in the top left panel of Figure~\ref{fig:1e657a} which makes use of 
a much deeper exposure. 
The inner edge is a cold front and the pressure across it is continuous 
within the errors, whilst the outer edge is a shock front with a large 
pressure jump consistent with a shock with Mach number $M=3$ 
\citep{markevitch2007,markevitch2002}. This is one of the best examples of a 
remnant type cold front, where the cool core of the merging subcluster survives
the initial infall, forming a cold front at its interface with the 
hot, shocked main ICM. The less dense outer subcluster ICM has been stripped 
from the denser core which survives because of its high density. This stripped 
gas can be seen in the X-ray image as a fan-like structure extending out behind 
the subcluster.

The top right panel of Figure~\ref{fig:1e657a} shows the temperature map with surface brightness 
contours overlaid. The temperature structure is complex showing pockets of hot 
($\sim20$\,keV) gas mixed with cooler ($\sim12$\,keV) gas. The only regular 
structure lies in the region containing the cold and shock fronts, where it can
be seen that the subcluster/cold front coincides with the region of 
$\sim7$\,keV gas (slightly hotter than the temperature measured above of 
$\sim5.8$\,keV due to projection effects), significantly cooler than the 
surrounding gas, whilst the shock front coincides with an arc of $18-20$\,keV 
gas. The hot regions on the eastern side of the main cluster are probably shock
heated regions of either stripped subcluster gas, or main cluster gas. Previous
temperature maps \citep{markevitch2002,govoni2004} also show 1ES0657-558 to have
a complex morphology with shock heated gas of up to 20\,keV, consistent with 
the map presented here. However, the temperature map we present here has higher 
resolution and statistical quality, since it is derived from almost an order of 
magnitude more \chan\, data. \citet{million2008} used the same data set to produce
a temperature map using an adaptive binning technique. The temperature map 
presented here agrees well with the \citeauthor{million2008} temperature map.

The cluster hosts a luminous radio halo \citep{liang2000}, and 
\citet{govoni2004} showed that the halo brightness peaks correspond to hot spots 
in their temperature map, and the halo extends along the shock front. 1ES0657-558
has been the subject of a thorough campaign to map its mass distribution using 
the combination of weak and strong lensing \citep{clowe2006,bradac2006}, the 
results of which have shown an offset between the lensing mass distributions and
the X-ray centroids providing compelling evidence for the existence of dark 
matter \citep{clowe2006}. These results, combined with the observations of 
\citet{barrena2002} and \citet{markevitch2002} show that the subcluster is 
probably caught just after its first (slightly off center) core passage, 
traveling from the south-east, and the merger occurs mostly in the plane of 
the sky, with the velocity difference between the main cluster and the 
subcluster only $\sim600$\kms \citep{barrena2002}. The velocity of the shock 
derived by \citet{markevitch2007} is 4700\kms, although in reality the 
subcluster is unlikely to be traveling at such a high velocity 
\citep[see, for example, ][]{springel2007,milosav2007}.

\subsubsection{Abell~3667}
\begin{figure*}
  \begin{center}
      {\includegraphics[angle=0,width=\textwidth]{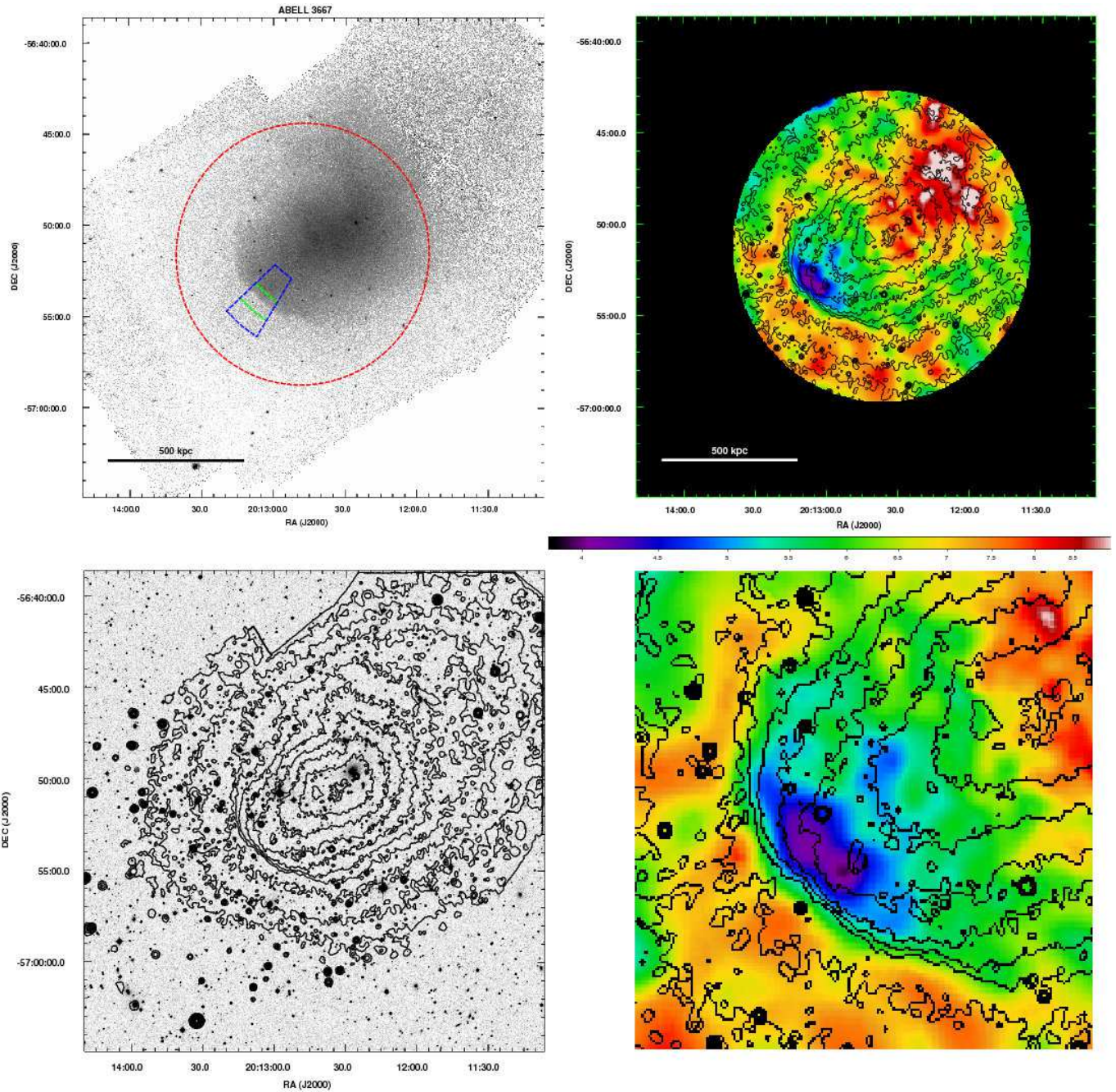}}
      \caption{Same as Figure~\ref{fig:1e657a}  but for Abell~3667 and with a Supercosmos Sky Survey $r_{\rm F}$-band
image at the lower left.}
      \label{fig:a3667a} 
    \end{center}
\end{figure*}

Abell~3667 was one of the original cold front clusters observed with \chan\, 
and, along with Abell~2142, provided the defining characteristics of cold fronts
\citep{vikhlinin2001b}. Abell~3667 has an X-ray luminosity of 
$L_{\rm X}(0.4 - 2.4\,{\rm keV}) = 5.1 \times 10^{44}\, {\rm erg\ s^{-1}}$ 
\citep{ebeling1996}, temperature $kT \sim 7$\,keV 
\citep{knopp1996,markevitch1998, vikhlinin2001b, briel2004}, Abell 
richness class 2 and a redshift of z=0.055. In the top left panel of 
Figure~\ref{fig:a3667a} the 
\chan\, X-ray image of the central core region of Abell~3667 is presented 
revealing a complex morphology which is elongated along a south-east to 
north-west direction with the cold front being the most prominent feature to the
south-east. The front appears regular at small opening angles, but fans out at 
larger angles to form a mushroom-cap-like shape. The same morphology is also 
prominent in the temperature map in the top right panel of Figure~\ref{fig:a3667a} 
which shows clearly 
that the front delineates regions of cool ($\sim 4.5$\,keV) gas from the 
ambient ICM with temperature ($\sim 7$\,keV). As noted by \citet{briel2004}, 
the gas at the tip of the front is coolest ($\sim 4$\,keV). \citet{heinz2003} 
use simulations to show motions inside a remnant gas core are induced due to 
the flow of gas around the cloud. These internal motions move lower entropy gas 
towards the tip of the cold front and, as the gas moves to the tip of the cold 
front, it cools due to adiabatic expansion, enhancing the temperature contrast 
at the tip of the front. The temperature map presented here is in good agreement
with those presented in the literature \citep{markevitch1999, vikhlinin2001b, 
mazzotta2002,briel2004}, although the longer combined exposure time of the 
\chan\, data presented here gives a more statistically significant map with 
higher spatial resolution. The map shows a complex morphology of hot and cool 
gas patches, including a patch of significantly hotter ($\sim 9 - 10$\,keV) 
gas to the north-west of the cluster center. 

At the cold front, a density jump of $2.6\pm 0.2$ is measured 
(see Table~\ref{front.props}). This 
value is inconsistent with the value of $3.9\pm0.8$ measured by 
\citet{vikhlinin2001b}. This is because \citeauthor{vikhlinin2001b}'s density 
measurement just outside the front is obtained by extrapolating
their $\beta$-model electron density model from larger radii to the front
position. Their aim was to determine the pressure in the undisturbed region of the
flow around the front, so they tried to correct for gas within the 
``compression'' region just outside the front where they note that the surface 
brightness $\sim 70$\,kpc exterior to the front exceeds that predicted by their 
$\beta$-model. For the density measurement made here, the power law density model 
describing the surface brightness outside the front is approximately limited to 
this $\sim 70$\,kpc region, meaning a higher density just outside the front is 
measured compared to that of \citeauthor{vikhlinin2001b}, which leads to a lower 
density jump.

Abell~3667 presents clear evidence for merger activity at both optical and radio
wavelengths. Observations at radio wavelengths show Abell~3667 is one of a 
handful of objects which harbor two diffuse radio relic sources 
\citep{rottgering1997}---one to the north-west and one to the south-east. The
radio observations also show two tailed radio galaxies, one of which lies on an
axis joining the two dominant cluster galaxies with its tail pointing along this
axis \citep{rottgering1997}. Radio relics are thought to trace merger shock 
fronts and \citet{roettiger1999} were able to reproduce the basic radio 
morphology of the relics using magnetohydrodynamic/N-body simulations of a 
slightly off-center merger with a 5-to-1 mass ratio and the subcluster impacting
from the south-east to the north-west, being observed $\sim 1$\,Gyr after core 
passage. The simulations also reproduced the basic observed X-ray morphology 
and galaxy distribution. 

Optical observations have shown the galaxy density distribution is bimodal, 
with the second density enhancement centered on the second brightest cluster 
galaxy \citep{proust1988,sodre1992,owers2009b}. Significant emission associated
with the secondary density enhancement was detected in the ROSAT X-ray 
observations of \citet{knopp1996}, lying just out of the field of 
view of the \chan\, observations. Using weak lensing, \citet{joffre2000} found evidence
for substructure in their mass maps which coincides with this secondary density 
enhancement. \citet{sodre1992} found the second D galaxy 
has a velocity offset of $\sim 400$\kms\, compared the the central D galaxy, 
and the velocity distribution surrounding this second D galaxy appears to 
indicate that it forms a dynamical subunit within the cluster. 

Abell~3667 was the subject of comprehensive optical follow-up with the aim of 
showing the cold front was related to merger activity in \citet{owers2009b}. 
Using combined spatial and velocity information for a large sample of 
spectroscopically confirmed cluster members, \citeauthor{owers2009b} found the 
cluster can be separated into two major components, coinciding with the density 
enhancements found by \cite{sodre1992} and separated in peculiar velocity by 
$\sim 500$\kms, plus a number of smaller subgroups. These observations indicate
Abell~3667 is undergoing a major merger which has produced the observed cold 
front.

\subsubsection{Abell~1201}
\begin{figure*}
  \begin{center}
      {\includegraphics[angle=0,width=\textwidth]{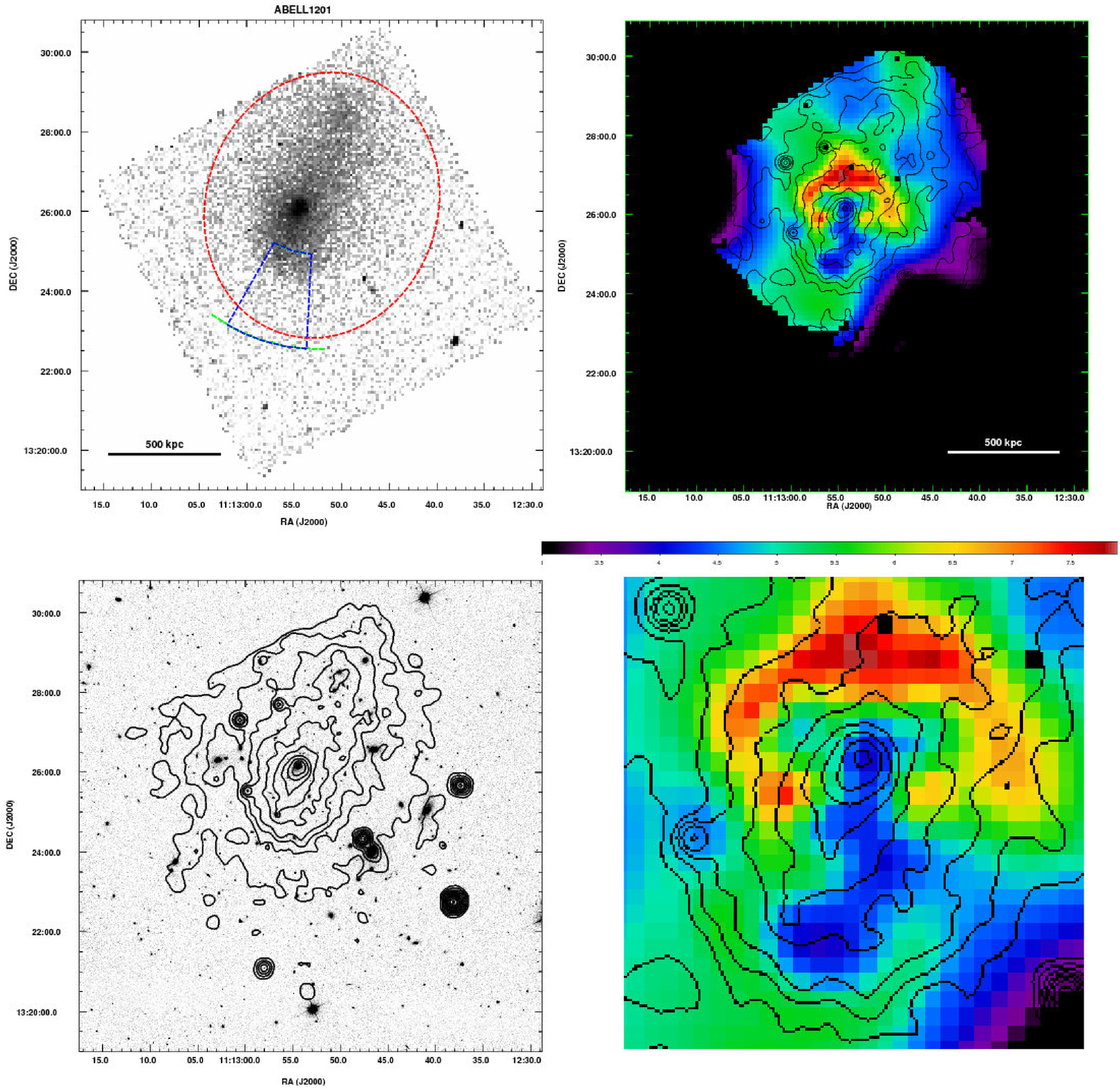}}
    \caption{Same as Figure~\ref{fig:1e657a} but for Abell~1201 and with an SDSS $r$-band image at the lower left.}
      \label{fig:a1201a}     
    \end{center}
\end{figure*}

Abell~1201 is one of the least well studied clusters in the sample and has an 
Abell richness class 2 \citep{abell1989}, redshift z=0.168 \citep{struble1999} 
with an X-ray luminosity $L_{\rm X}(0.1-2.4\,{\rm keV})=2.4~\times~10^{44}\,\rm{ergs}\, \rm{s}^{-1}$ 
\citep{bohringer2000}. The top left panel of Figure~\ref{fig:a1201a} shows the 
\chan\, image of 
Abell~1201 which is elongated along a south-east to north-west axis, with the 
main cold front clearly visible to the south-east, a bright central core which 
also has a cold front towards the north-west and a remnant core also visible to
the north-west. The temperature map shown in the top right panel of 
Figure~\ref{fig:a1201a} reveals a 
hot $\sim7$\,keV region lying between the main cluster core and the remnant 
core, indicating some form of heating here probably caused by adiabatic 
compression due to core motion. There is no evidence for significant temperature
differences at the location of the remnant core (c.f. \bulletclus\ where the remnant
core clearly stands out in the temperature map). This may be attributed to the 
heavy smoothing required to obtain well-defined temperatures, since this 
observation was significantly affected by strong flares, so that nearly half of 
the exposure was rejected during cleaning. There is a cool ($\sim4$\,keV) 
stream of gas connecting the cold front to the cool core of the main cluster in
the temperature map. This indicates that the cold gas associated with the front
is gas which has been displaced from the cool core of the cluster, and the 
front is probably associated with a spiral-like structure (for example as seen in 
the temperature maps for MS1455.0+2232 and \rxj\ in the top right panels of  
Figures~\ref{fig:ms1455a} and \ref{fig:rxj1720a}, respectively) seen in projection, 
similar to those seen in the $x$ projection image presented in Figure~19 of 
\citet{ascasibar2006}. 

Detailed follow up multi-object optical spectroscopy of Abell~1201 has been 
undertaken and is presented in \citet{owers2009a}. This study shows a bimodal
galaxy density distribution with the main component centered on the main X-ray
core and a second component coincident with the remnant core to the north-west. The 
combination of spatial and velocity information also reveals the presence of 
substructure coincident with the remnant core and offset in peculiar velocity by 
$\sim 430$\kms, indicating a merger is occurring with the majority of its 
velocity in the plane of the sky. The perturbation from this merger has
given rise to the cold fronts caused by sloshing of the ICM. 

\citet{edge2003} discovered a small scale gravitational arc in Abell~1201, and 
proposed the observed arc properties could be explained by a mass distribution 
which was either highly elliptical or bimodal. The data allowed a 
bimodal model with a secondary mass clump lying to the south-east of the cluster 
center, however, \citeauthor{edge2003} favored the model with high ellipticity.
Given the results of \citet{owers2009a}, it would seem that a bimodal mass distribution
is the most likely explanation for the data, although the center of mass of the 
second component is coincident with the north-west remnant core.

\subsubsection{Abell~2069}
\begin{figure*}
  \begin{center}
    {\includegraphics[angle=0,width=\textwidth]{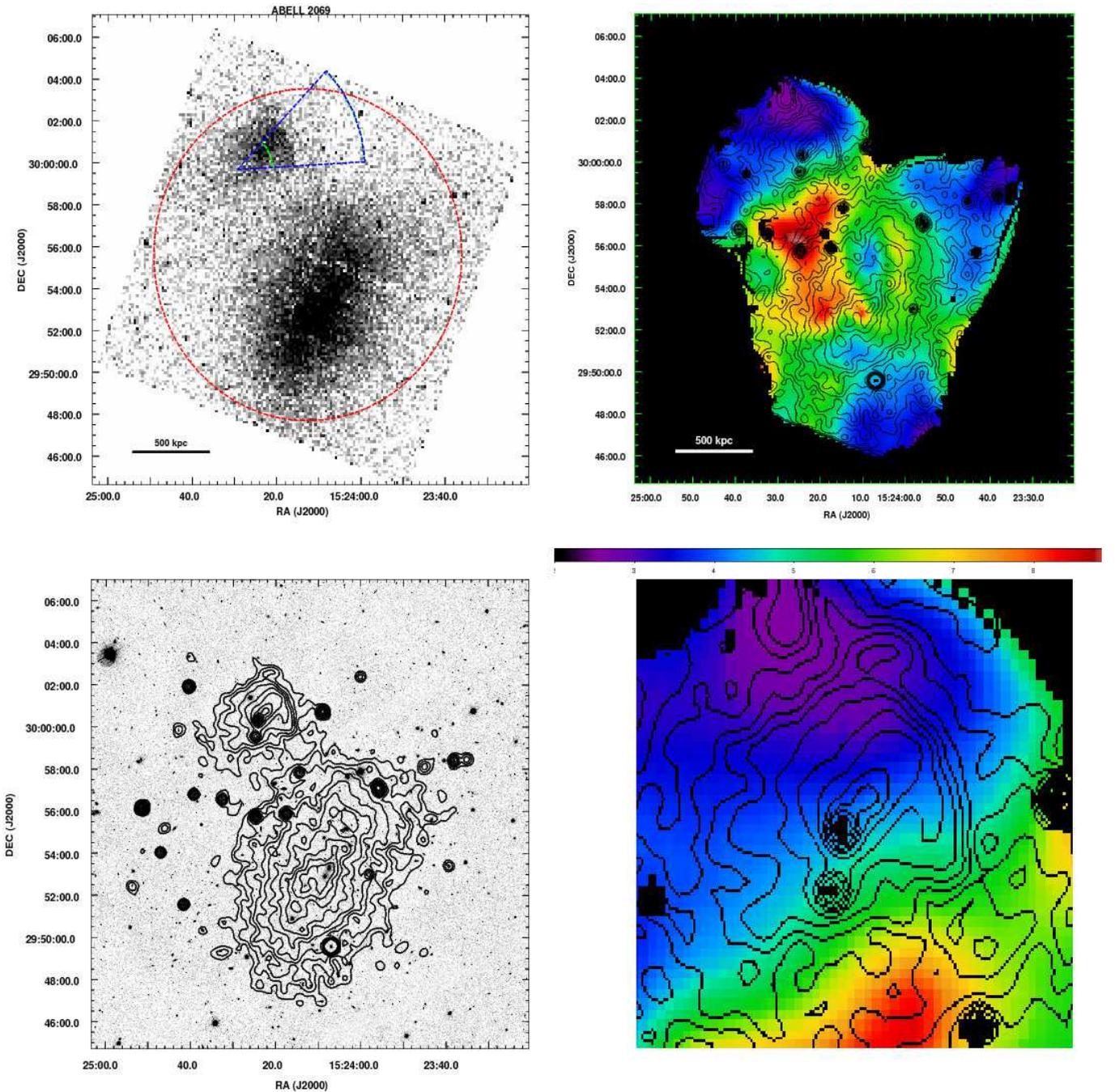}}
    \caption{Same as Figure~\ref{fig:1e657a} but for Abell~2069 and with an SDSS $r$-band image at the lower left.}
      \label{fig:a2069a} 
    \end{center}
\end{figure*}

Abell~2069 is a richness class 2 \citep{abell1989} cluster at redshift z=0.116, 
which was originally shown to have multiple condensations in the gas and galaxy
distributions by \citet{gioia1982} and is part of a large supercluster 
containing some 6 clusters \citep{einasto1997}. The top left panel of 
Figure~\ref{fig:a2069a} shows the 
\chan\, image of Abell~2069 (binned to $7.8''$ pixels) which shows the main 
component \citep[Abell~2069A in the terminology of ][]{gioia1982} to be elongated 
along a south-east to north-west axis. There are two bright elliptical galaxies
aligned along this axis which differ in peculiar velocity by only
$\sim 230$\kms\,\citep{gioia1982}. Abell~2069B can be seen to the north-east and
it hosts a cold front towards the north-west along with a fan like extension of
X-ray emission towards the south-east. There is a dominant elliptical galaxy 
associated with Abell~2069B which has a redshift z=0.1178 and is surrounded by 
an overdensity in the projected galaxy number surface density \citep{gioia1982}.
Assuming the redshift of the dominant elliptical is representative of 
Abell~2069B, then its peculiar velocity with respect to the mean cluster 
redshift is only $485$\,\kms. Although the cold front is associated with the 
Abell~2069B, it is difficult to determine whether the nature of this 
cold front is consistent with a contact discontinuity occurring at the interface
of the lower entropy Abell~2069B ICM and the higher entropy Abell~2069A ICM, or
a cold front induced by sloshing of the Abell~2069B gas. Given the small 
peculiar velocity offset of Abell~2069B and Abell~2069A, it can be concluded 
that the majority of the subcluster motion is in the plane of the sky. 

The resolution of the temperature map presented in the top right panel of 
Figure~\ref{fig:a2069a} is 
limited by a lack of photons. However, it does show that the projected 
temperature around Abell~2069B is significantly cooler than the ICM surrounding
it, and also that Abell~2069A hosts a complex temperature distribution with 
regions of $6-8$\,keV gas interspersed with pockets of cooler $\sim4$\,keV gas.
The temperature structure in Abell~2069A may be due to a core crossing by 
Abell~2069B. In particular the region of $6-8$\,keV gas to the east may be due 
to the north-east bound passage of  Abell~2069B. In this case, the cold front in
Abell~2069B is more likely to be due to sloshing of the Abell~2069B gas.
However, the existence of the
two dominant elliptical galaxies may be indicative of merger activity within 
Abell~2069A itself and this could also explain the complex temperature 
structure.

\subsubsection{Abell~1758N}
\begin{figure*}
  \begin{center}
      {\includegraphics[angle=0,width=\textwidth]{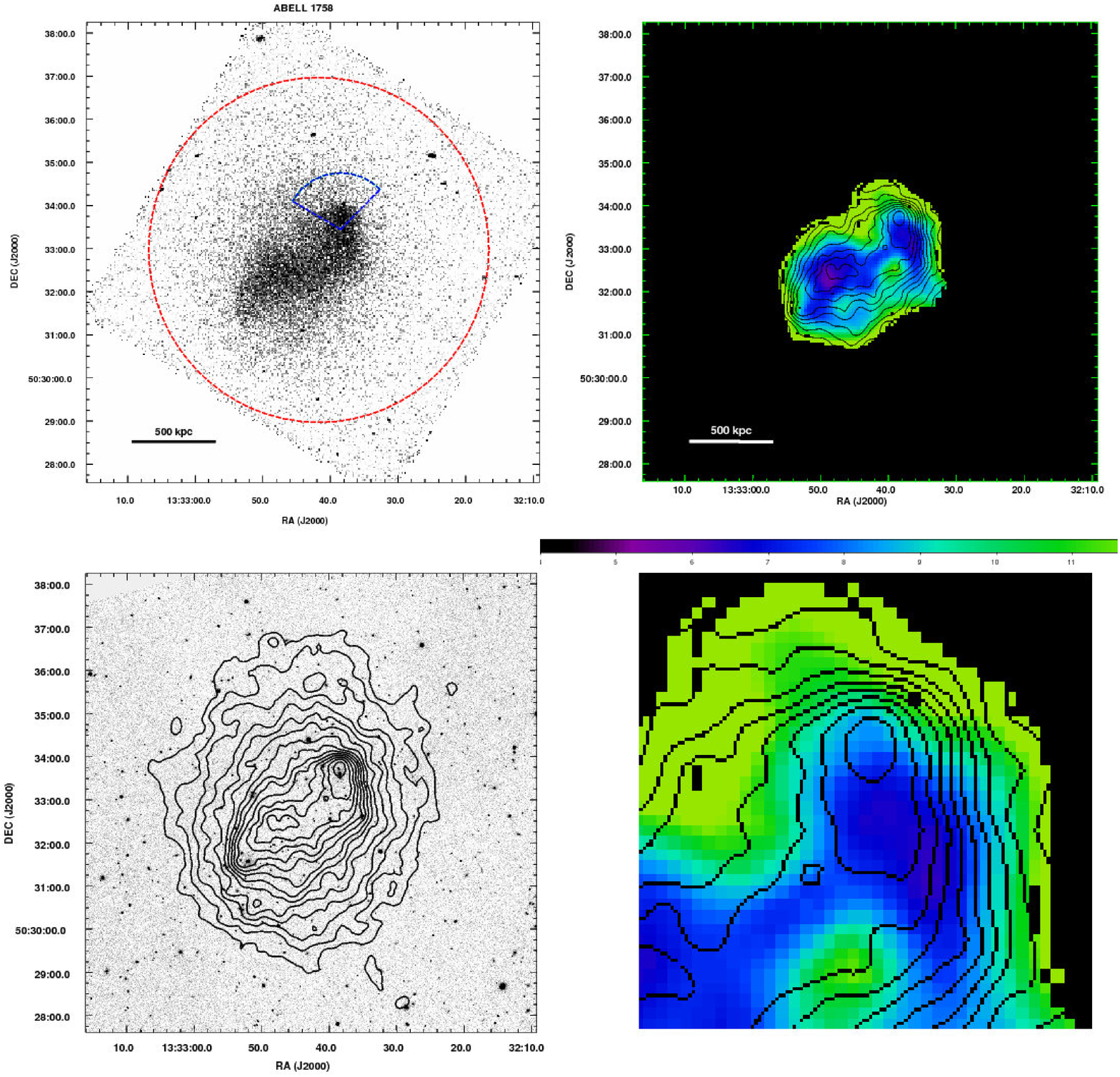}}
    \caption{Same as Figure~\ref{fig:1e657a} but for Abell~1758 and with an SDSS $r$-band image at the lower left.}
      \label{fig:a1758a} 
    \end{center}
\end{figure*}

Abell~1758 is a rich \citep[Abell richness class 3;][]{abell1989} double cluster
at redshift z=0.279 \citep{struble1999}. The \chan\, observations, which were 
originally presented in \citet{david2004}, were taken on the ACIS-S3 chip and 
contain only the northern component of the double cluster which is referred to 
as Abell~1758N. The \citet{david2004} study showed the X-ray 
morphology is irregular with two main components placed on an axis running from 
the south-east to the north-west. The \cite{david2004} study also showed that the 
morphology of the north-west component is fan-like with a tail of gas extending 
to the south, and a cold front located towards the north, suggesting the motion 
of the substructure is towards the north. These features can be seen in 
top left panel of Figure~\ref{fig:a1758a} where \chan\ X-ray image is reproduced. Using a truncated 
spheroid to model the surface brightness across the front, \citet{david2004} 
measured a density jump of $1.6\pm0.2$ ($1\sigma$ errors) which is consistent 
within the errors with the value presented in Table~\ref{front.props}. The 
north-west and south-east structures are coincident with peaks in the projected 
galaxy number density maps presented in \citet{dahle2002} and also with 
substructure in the lensing maps of \citet{okabe2008}. There is a second less 
significant front further south-east of the south-east structure for which 
\citet{david2004} measured a density jump of $\sim 1.5\pm0.2$. They used a 
hardness ratio map to show that the gas inside the front is cooler than outside, 
thus concluding it is also a cold front caused by the interface of low entropy 
subcluster gas with the higher entropy ambient ICM. Coincident with the parcel 
of low entropy gas responsible for the south-east cold front is another 
overdensity of galaxies housing a bright elliptical galaxy.

The top right panel of Figure~\ref{fig:a1758a} shows the temperature map for Abell~1758N, which only 
covers a small area due to the low number of photons detected, 
particularly in the outskirts. The map shows the two components host 
significantly cooler gas  compared to the hotter gas in the 
outskirts. The absence of a sharp temperature jump at the cold front is mainly
due to the poor effective resolution of the temperature map. This means the cold 
front regions are contaminated by the hotter gas from the outskirts, thus the 
temperature measured here is the average of the two contributions. There is a 
bridge of cool gas joining the two subclusters, and a tongue of cool gas trailing 
the south-east subcluster which is presumably ram pressure stripped gas due to the 
collision of these two subclusters. The features seen in the temperature map are
qualitatively similar to those seen in the hardness ratio map of \citet{david2004}.

Based on the X-ray characteristics, \citet{david2004} presented a merger scenario
for Abell~1758N whereby the system is being viewed after a recent off-axis 
pericentric passage of two $\sim 7$\,keV clusters, and the significantly hotter
gas in the outskirts has been heated by merger shocks. Indirect evidence for 
merger activity occurs at radio wavelengths where, at the position of the 
south-east cold front, there is a narrow-angle tailed (NAT) radio galaxy which 
has its tail extending from the north-west to the south-east \citep{odea1985}, 
and its head coincident with a fainter galaxy to the north-east of the dominant
south-east elliptical.  \citet{bliton1998} showed that NAT galaxies are 
preferentially found in clusters with substructure, concluding that at least 
part of the relative velocities required to produce a NAT is due to bulk gas
motions generated during a merger. Further \chan\, observations with longer 
exposures would enable the measurement of pressure differentials at the cold 
front, thus constraining gas velocities and allowing the components of the 
relative velocities of the gas and NAT galaxy to be constrained. 

Given the fan-like morphology and its coincidence with a galaxy overdensity, it
appears the north-west cold front in Abell~1758N is a merger induced cold front
generated at the interface of colliding subclusters.

\subsubsection{Abell~2163}
\begin{figure*}
  \begin{center}
      {\includegraphics[angle=0,width=\textwidth]{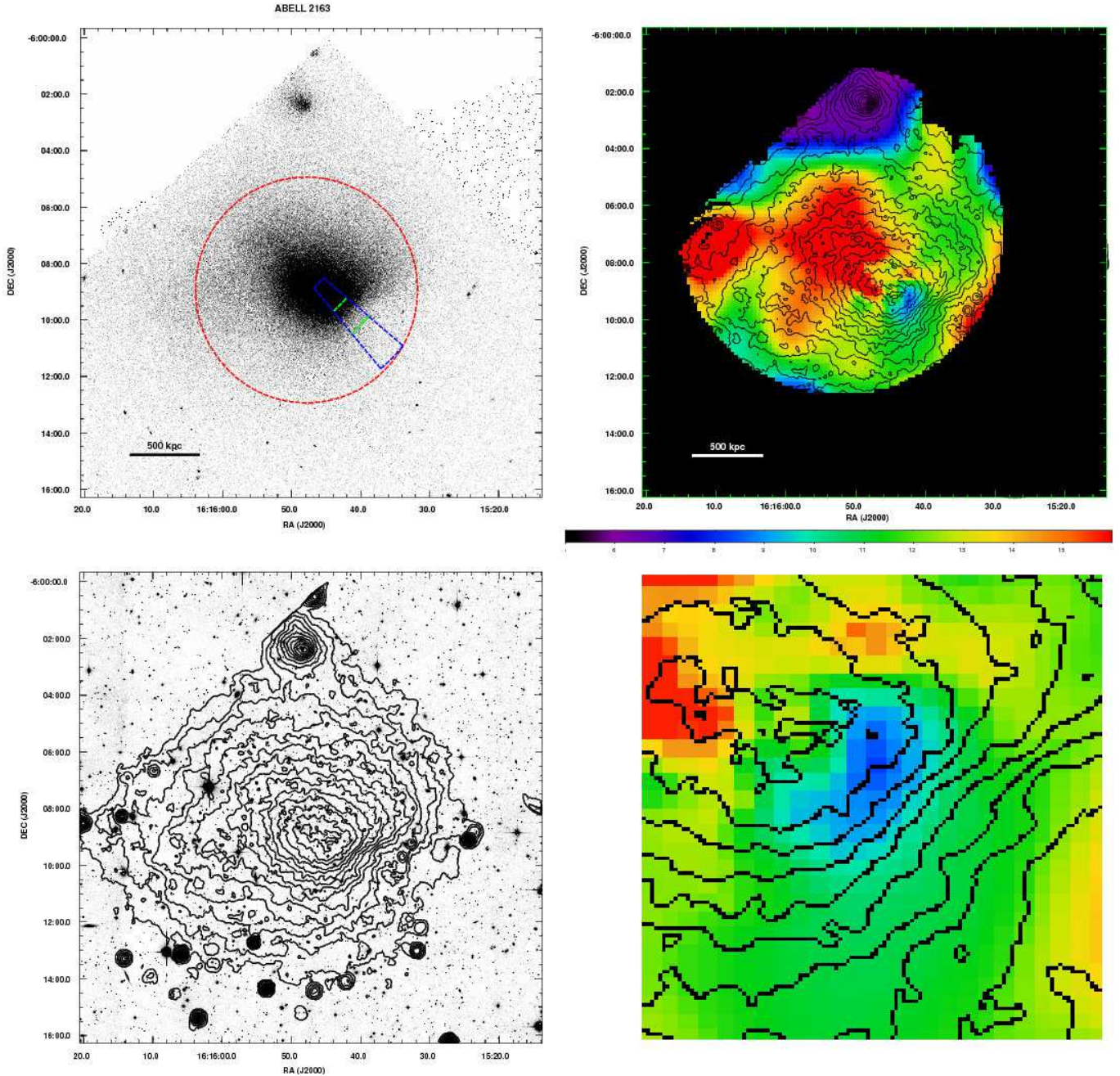}}
    \caption{Same as Figure~\ref{fig:1e657a} but for Abell~2163 and with a CFHT $r$-band MegaCam image at the lower left.}
      \label{fig:a2163a} 
    \end{center}
\end{figure*}

Abell~2163 is one of the hottest \citep[kT = 15.5\,keV;][]{maughan2008}, most 
luminous ($L_{\rm X}[0.1-2.4\,{\rm keV}]=22.9\times 10^{44}\,{\rm erg\ s^{-1}}$;
\citealt{ebeling1996}) clusters of galaxies known. It has a redshift of 
$z=0.201$, Abell richness class 2 \citep{abell1989} and has been the subject of 
many studies at multiple wavelengths \citep[e.g.;][]{maurogordato2008,
feretti2004,markevitch2001a,feretti2001,squires1997,holzapfel1997}. The 
\citet{markevitch2001a} \chan\ observations of Abell~2163 revealed a cluster 
with an irregular morphology 
with an extension along a direction going north-east to south-west, as seen in
deeper \chan\ image presented in top left panel of Figure~\ref{fig:a2163a}. There 
is a cold front in the south-west which appears to have a mildly convex shape and 
is not as sharp as others seen in this sample (e.g. the cold front seen in 
Abell~3667), possibly indicating the front has 
significant inclination with respect to the plane of the sky or is in the 
process of breaking up. There is a group due north and a hint of excess emission
east of the core emission. The temperature map shown in the top right panel of 
Figure~\ref{fig:a2163a} shows the cluster exhibits a complex multi-temperature gas 
distribution with the
cold front associated with $\sim 8$\,keV gas and surrounded by hotter 
$\sim 11$\,keV gas, consistent with the temperature jump given in 
Table~\ref{front.props} within the errors. The regions corresponding to the 
X-ray extension towards the south-east and the excess X-ray emission
to the east are significantly hotter than the rest of the cluster at $>14$\,keV.
The infalling group to the north is significantly cooler than the main cluster 
at 5-6\,keV according to the temperature map. The temperature map presented here
is in good agreement with the maps published by \citet{govoni2004} and 
\citet{markevitch2001}, although we obtain higher resolution and statistical 
quality by simultaneously fitting the two \chan\ data sets to produce the 
temperature map.

As mentioned above, Abell~2163 has been studied extensively at multiple 
wavelengths and evidence for merger activity presents itself in the majority of
these observations. The first indications that Abell~2163 was an atypical 
cluster came from combined {\it Ginga} and {\it Einstein} X-ray observations 
which showed an extremely high temperature \citep{arnaud1992}. Further ROSAT and
ASCA observations confirmed the high temperature and also showed evidence for 
temperature and surface brightness structure in the ICM, suggesting evidence 
for merger activity \citep{markevitch1994,elbaz1995,markevitch1996}. High 
resolution \chan\, temperature maps presented in \citet{markevitch2001} and 
\citet{govoni2004} revealed this temperature structure in greater detail, 
showing in particular that the central region harbors complex temperature 
structure which is consistent with a major merger, along with the 
group towards the north seen in the ROSAT images. \citet{squires1997}
shed light on the dynamical state of the cluster showing the distributions of 
both mass and galaxies are bimodal in the center, with the peaks in each 
distribution roughly coincident, and concluded the cluster is not in a relaxed 
dynamical state. At radio wavelengths, Abell~2163 hosts a luminous radio halo, 
a possible radio relic, along with 3 tailed radio galaxies all with tails 
oriented in the same direction towards the west \citep{feretti2001}. Finally, 
\citet{maurogordato2008} present an analysis of the cluster dynamics from a 
sample of 361 cluster member redshifts which show a multi-modal velocity 
distribution, strong velocity gradients along a north-east to south-west axis 
along with multiple clumps in the galaxy surface density. These results were
confirmed by \cite{owers2008} where the analysis of a sample of 491 cluster 
member redshifts showed a significantly skewed velocity distribution and 
significant dynamical substructure when spatial and velocity information was
combined. In particular, substructure was found to coincide with the cold front
seen in Figure~\ref{fig:a2163a}, strongly supporting the
argument that this cold front is caused by an ongoing cluster merger.

\subsection{Clusters with relaxed Appearing X-ray Morphologies}\label{relaxed}

\subsubsection{RXJ1720.1+2638}\label{rxj}
\begin{figure*}
  \begin{center}
      {\includegraphics[angle=0,width=\textwidth]{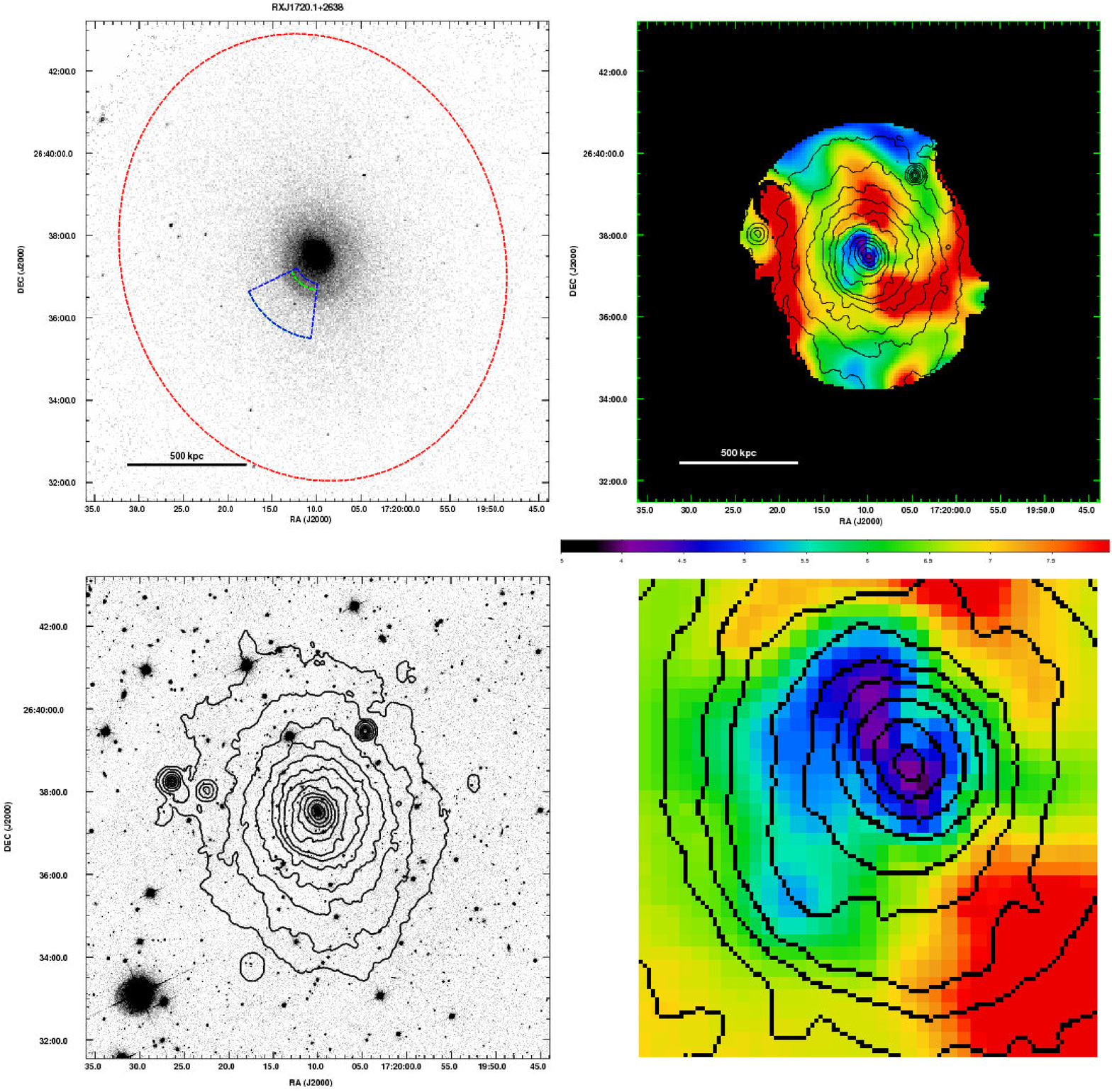}}
    \caption{Same as Figure~\ref{fig:1e657a} but for \rxj\ and with an SDSS $r$-band image at the lower left.}
      \label{fig:rxj1720a} 
    \end{center}
\end{figure*}

RXJ1720.1+2638 was the first of the otherwise relaxed appearing clusters which 
were revealed to have cold fronts by \chan\, \citep{mazzotta2001}. It has an 
X-ray luminosity of $L_{\rm X}(0.1 - 2.4\,{\rm keV})=7.3\times10^{44}\rm erg\ 
s^{-1}$ \citep{bohringer2000} and redshift  $z=0.1605$ \citep{owers2008}. 
The top left panel of Figure~\ref{fig:rxj1720a} shows the \chan\, X-ray image of 
RXJ1720.1+2638 which
reveals a relaxed overall morphology with a bright core and a cold front towards
the south-east. \citet{mazzotta2008} also report a cold front towards the 
north-west of the core which has a density jump of $\sim 1.68$. It is noted that
\citet{mazzotta2008} measured a density jump of $1.66\pm0.05$ for the south-east
cold front, inconsistent with the value given in Table~\ref{front.props}. This 
is because \citet{mazzotta2008} measured the density jump over a larger opening 
angle, which included regions where the front is less well defined. 
The temperature map presented in the top right panel of Figure~\ref{fig:rxj1720a} 
reveals more about 
the nature of the cold front. There is a finger of cool gas starting at the 
surface brightness peak and spiraling in an anticlockwise direction before 
terminating at the cold front, also seen in the temperature maps of 
\citet{mazzotta2008}, and similar to the temperature structure seen in the 
simulations of \citet[][see their Figure~7]{ascasibar2006}. On larger scales 
the temperature map is patchy with regions of hot $\sim 8$\,keV gas surrounded 
by ambient $\sim 6$\,keV gas.

\citet{dahle2002} report on the distribution of galaxies, light and dark matter
within RXJ1720.1+2638, finding the light distribution to be fairly regular and 
dominated by the central cluster galaxy. The weak lensing mass distribution 
appears to be less regular, with the main signal roughly coincident with the 
cluster center and a secondary peak detected $\sim 2 - 3$\,arcmin to the 
south-west. The galaxy number density also appears to be fairly regular, 
although there is an extension of the contours towards the south-west in the 
direction of the secondary mass clump detected in the weak lensing maps. 
RXJ1720.1+2638 is by all accounts a fairly relaxed looking system, and, along
with MS1455.0+2232, is thus 
a very important test case within our sample, given the hypothesis that cold 
fronts are a result of major merger activity. For this reason, RXJ1720.1+2638 
was selected for follow up observations using multi-object spectroscopy. The 
results of the study are presented in \citet{owers2008} and show evidence for 
substructure in the galaxy density distribution and also when spatial and 
velocity information is combined. However, the substructure's relation to the
cold front remains ambiguous, and further observations are required.

\subsubsection{MS1455.0+2232}\label{ms1455}
\begin{figure*}
  \begin{center}
      {\includegraphics[angle=0,width=\textwidth]{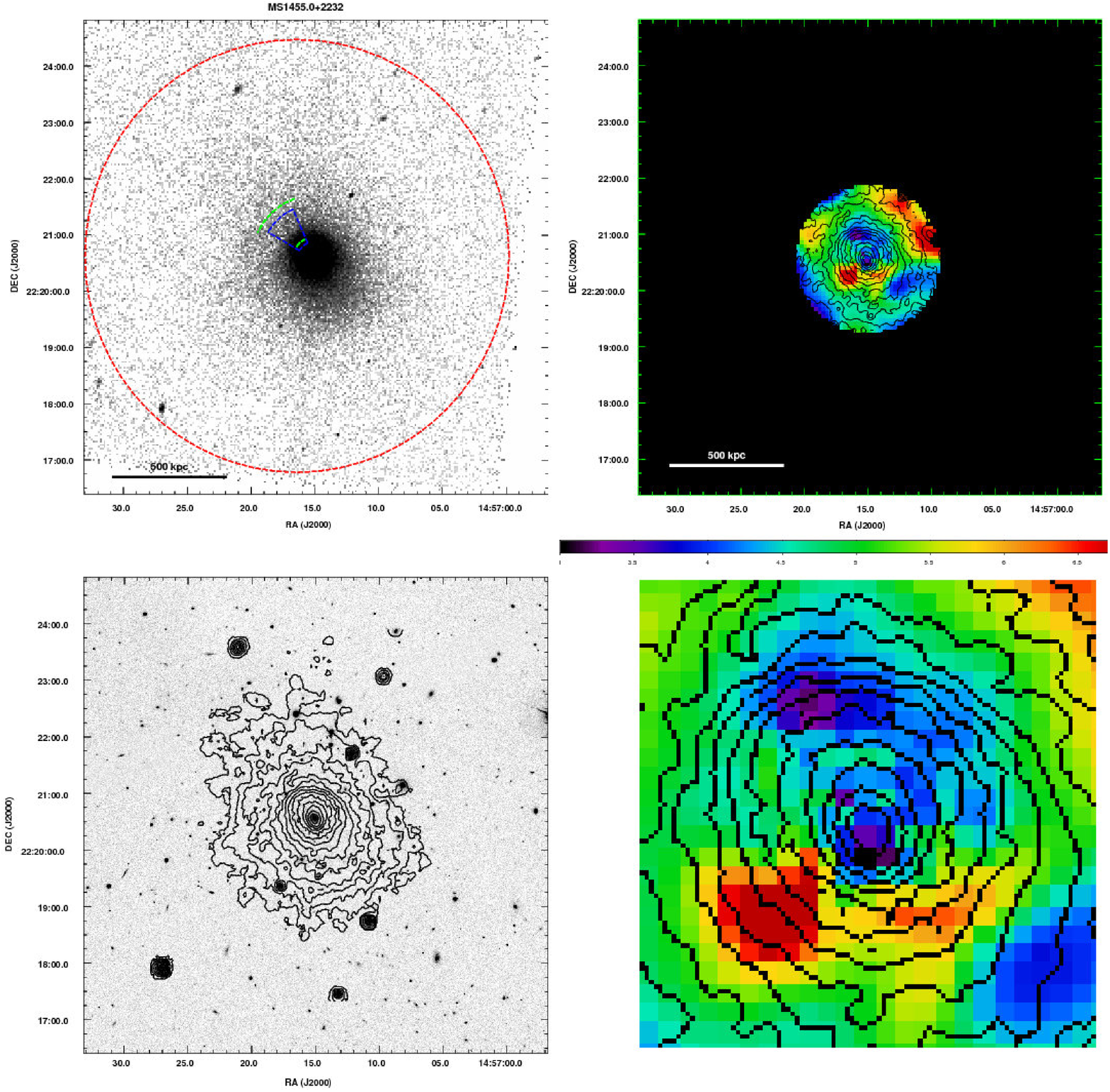}}
    \caption{Same as Figure~\ref{fig:1e657a} but for MS1455.0+2232 and with an SDSS $r$-band image at the lower left.}
      \label{fig:ms1455a} 
    \end{center}
\end{figure*}

MS1455.0+2232 (also known as ZW7146) was discovered serendipitously as an 
extended X-ray source in the {\it Einstein Observatory} Extended 
Medium--Sensitivity Survey \citep[EMSS;][]{gioia1990} and later reported to 
host a massive 1500\msolar$yr^{-1}$ cooling flow \citep{allen1996}. 
The top left panel of Figure~\ref{fig:ms1455a} shows the \chan\, X-ray image of 
MS1455.0+2232 which 
has a relaxed X-ray morphology, harboring a slight ellipticity with major axis
running from the south-west to north-east with a mild asymmetry along this axis.
Also along this axis is the cold front as outlined by the blue annular segment 
in the top left panel of Figure~\ref{fig:ms1455a}. Inspection of the temperature 
map presented in the top right panel of Figure~\ref{fig:ms1455a} reveals a 
spiral-like feature of cool gas beginning at
the cool core and ending at the position of the cold front. This feature is also 
seen in the temperature map of \citet{mazzotta2008}, which is derived from the 
same \chan\ data set, and shows good overall agreement with the map presented here. 
The cold front appears similar to those seen in the simulations of  
\citet{ascasibar2006} (see Figure 7 and Figure 19) where an off-center merger 
of a gasless dark matter halo disturbs the cool cluster core, offsetting both 
the gas and dark matter from its initial position. The offset low entropy gas 
does not fall back into the core along a purely radial path, since it has 
acquired angular momentum from the infalling dark matter halo, and thus a 
spiral pattern is formed.

Further evidence for merger activity in MS1455.0+2232 comes from the weak 
lensing maps of \citet{dahle2002}, where, despite showing a fairly regular 
projected galaxy number density distribution, the weak lensing maps showed a 
highly elliptical morphology with major axis pointing along an axis going from 
south-west to north-east (similar to the X-ray major axis angle). 
\citet{smail1995} found the position angle for the  mass, gas and projected 
galaxy density distributions within $\sim 1.25$ arcmin ($\sim 300$\,kpc) are all
aligned along the same axis. The combined radio and X-ray observations of 
\citet{mazzotta2008} have shown a spatial correlation between the spiral 
structure and radio emission associated with a core mini radio halo, suggesting
a population of relic electrons injected by the central AGN have been 
reaccelerated by turbulence associated with the gas motions causing the cold 
fronts. 

\subsubsection{Abell~2142}
\begin{figure*}
  \begin{center}
      {\includegraphics[angle=0,width=\textwidth]{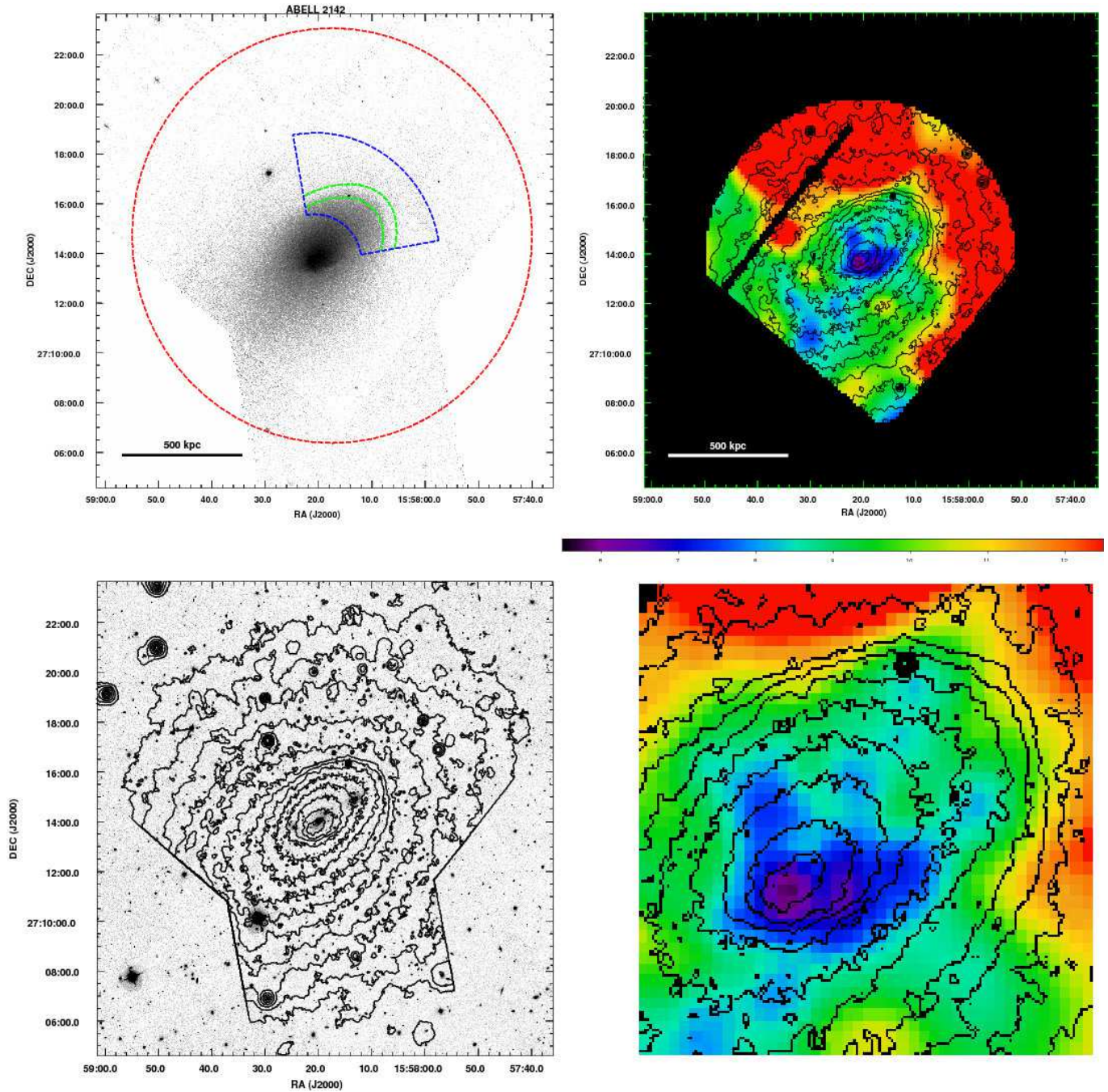}}
    \caption{Same as Figure~\ref{fig:1e657a} but for Abell~2142 and with an SDSS $r$-band image at the lower left.}
      \label{fig:a2142a}    
    \end{center}
\end{figure*}

The first cold fronts were discovered in \chan\, observations of Abell~2142
\citep{markevitch2000}. The cluster is optically rich \citep[Abell richness
class 2;][]{abell1989}, hot \citep[kT=9\,keV;][]{white1994,allen1998,markevitch1998,markevitch2000}, X-ray luminous 
($L_{\rm X} [0.1 - 2.4\,{\rm keV}] = 11.4\times 10^{44}\rm erg\ s^{-1}$; 
\citealt{bohringer2000}) with redshift z=0.089 and velocity dispersion 
$\sigma = 1280$\kms\, \citep{oegerle1995}. The top left panel of 
Figure~\ref{fig:a2142a} shows the 
\chan\, X-ray image of Abell~2142 which has an elliptical morphology with its 
position angle aligned along a south-east to north-west axis. Along this axis 
and to the north-west is the cold front for which density and temperature jumps
were measured, whilst a second less prominent cold front, noted by 
\citet{markevitch2000}, lies just south of the
core. The north-west cold front appears to have a cometary morphology with
the sharp brightness edge to the north-west forming the head, and a 
tail pointing towards the south-east where the surface brightness drops less 
sharply with radius. The temperature map presented in the top right panel of 
Figure~\ref{fig:a2142a} 
is generated using a much longer exposure (ObsID 5005 data) than the one shown 
in \citet{markevitch2000} and reveals a complex temperature distribution, where 
the two cold fronts clearly 
separate regions of cool and hot gas. The temperature morphology of the
central region is very similar to that seen by \citet{markevitch2000}, i.e.,
the brightness peak is much cooler ($\sim 6$\,keV) than the gas at larger radii,
and there is a finger of $\sim7$\,keV gas extending from the core towards the 
north-west. The regions inside the north-west cold front are hotter than the 
core gas, at $\sim 8 - 9$\,keV, but significantly cooler than the gas at larger 
radii outside the front ($>12$\,keV). Note that towards the south-east, 
the gas temperature is around $7 - 8$\,keV, inconsistent with the rise 
in temperature seen in the same region by \citet{markevitch2000}. This 
difference is likely attributed to the different data 
sets used---\cite{markevitch2000} used early ACIS-S data where the calibration
was poorly known, whereas here the ACIS-I data are used along with the latest 
and much improved calibration files.

The exact nature of the cold fronts is uncertain. \citet{markevitch2000} 
propose they are due to the survival of dense subcluster cores during a major 
merger, but later \citet{markevitch2007} claim the cold fronts are more likely 
to be the result of core gas motion (not unlike RXJ1720.1+2638, MS1455.0+2232 
and Abell~1201 of this sample). However, Abell~2142 seems anomalous since, while
there is a finger of cool gas extending from the core towards the north-west, 
this finger does not form a coherent spiral-like temperature structure joining 
the core to the north-western cold front, as seen in the clusters MS1455.0+2232 
and RXJ1720.1+2638. This of course could be due simply to different viewing 
angles, since the existence of multiple fronts lying on different sides of the 
cluster core is fairly compelling evidence for sloshing type cold fronts.

Power ratio analyses of ROSAT X-ray images gave an indication that Abell~2142 
was a dynamically active system \citep{buote1996}, and the case for a merger was
strengthened by the analysis of \citet{henry1996}, which showed strong 
variations in the azimuthal temperature distribution. In an earlier study using
103 cluster member redshifts, \citet{oegerle1995} found evidence for 
substructure at the $90\%$ confidence level using the Dressler-Shectmann 
$\Delta$ test, although they were unable to isolate the substructure spatially 
or in velocity space and conclude Abell~2142 is a cluster with uncertain 
dynamics. Indirect evidence for dynamical activity came from the observation of 
two dominant bright elliptical galaxies in the cluster core, with one having a 
redshift consistent with the cluster mean redshift and the second offset in 
peculiar velocity by $\sim 1600$\kms. Further evidence for dynamical activity 
came from the weak lensing maps of \citet{okabe2008}, who found evidence for a 
significant excess in their weak lensing signal at a projected distance of 
$\sim540$\,kpc ahead of the north-west cold front. At radio wavelengths, 
Abell~2142 exhibits diffuse radio emission extending $\sim 350$\.kpc about the 
cluster core, although it is unclear whether this is part of a massive radio 
halo associated with a cluster merger \citep{giovannini1999}. \citet{bliton1998}
show a NAT source which has a tail extending away from the cluster center and in
roughly the same direction as the major axis of the X-ray emission. This NAT 
galaxy is displaced by $\sim1400$\kms\, from the cluster rest frame, and its 
head is located just within the edge of the north-west cold front (coincident 
with the X-ray point source seen to lie just within the north-west cold front in
Figure~\ref{fig:a2142a}). Despite the uncertain nature of the merger geometry, 
it is clear that Abell~2142 is a dynamically active cluster. 

\section{Discussion}\label{discussion}

Selecting clusters based purely on the existence of a cold front has produced a 
sample with two distinct X-ray morphological types: those with clearly disturbed X-ray 
morphologies (Section~\ref{disturbed}) and those with fairly relaxed X-ray 
appearances, aside from the cold fronts (Section~\ref{relaxed}). In 
Table~\ref{merger.evidence}, the evidence indicating merger activity for the 
cold front sample, along with the X-ray morphologies, is summarized. Once again,
to emphasize the dichotomy, the clusters in Table~\ref{merger.evidence} are
separated into the two subsamples presented in Section~\ref{cf.clusters}.

Of the six clusters in Section~\ref{disturbed}, five have significant evidence 
for merger-related activity at other wavelengths and, as was evident from their 
X-ray morphologies, it is clear the existence of a cold front is good evidence 
for merger activity\footnote{Note that the existence of additional structure 
at other wavelengths was not considered in the selection of the 
cold front sample.}. The sixth cluster, Abell~2069, does have evidence for 
substructure in its projected galaxy density \citet{gioia1982}, although it is 
not as well studied at multiple wavelengths when compared to the other five 
clusters, thus the evidence for merger-related activity at other wavelengths 
is less compelling. Within this subsample there are a broad range of X-ray 
morphologies. This can 
be traced to the different formation mechanisms for 
cold fronts which depend on the details of the merger event \citep[e.g. merger 
stage, mass of subcluster, impact parameter or the existence of cool cores;][]{poole2006}. 
For example, upon comparing Abell~1201 to \bulletclus, it can be seen that both
contain remnant cores observed after pericentric passage. However, Abell~1201's cold
front is due to gas sloshing of the primary's cool core, while the cold front in  
\bulletclus\ occurs in the remnant core and there is no evidence for a cold front 
due to motion of its primary's core. Two scenarios relating the details of 
the initial conditions of the mergers to the cold front formation 
can explain this: (1) The primary in 
\bulletclus\ did not contain a significant cool core, thus no cold fronts 
analogous to those in Abell~1201 are seen. (2) The impact parameters of the 
mergers differ significantly, i.e., the merger in the \bulletclus\ was almost 
directly head on, thus the primary's core was destroyed by the secondary's 
passage, while the pericentric passage of the secondary in Abell~1201 was 
off-axis and did not significantly disrupt the cool core allowing the formation of
the cold fronts. Thus, aside from being good merger signposts, cold fronts also 
provide additional information which can be used in tandem with observations
at other wavelengths to constrain the merger histories in these clusters.

Turning to the clusters in Section~\ref{relaxed}, Table~\ref{merger.evidence}
shows that currently there is very little observational evidence in the literature 
to determine a clear cut explanation for the cold fronts in clusters with overall 
relaxed morphologies. The simulations of \citet{ascasibar2006} showed that pure 
gravitational disturbances in the form of infalling gasless dark matter subclumps 
are able to produce these cold fronts whilst maintaining the overall relaxed 
appearance. Commonly seen in simulations of this scenario are spiral-like 
temperature structures, which we also see in the temperature maps of RXJ1720.1+2638 
and MS1455.0+2232 (Figures \ref{fig:rxj1720a} and \ref{fig:ms1455a}, respectively), 
showing the cool gas causing the front to be connected to the cool cluster core.
Also apparent in these simulations are the existence of multiple cold fronts at
different radii; a signature clearly observed in Abell~2142. 
Marginal evidence is found favoring this scenario in \rxj\, \citep{owers2008}, 
where there is substructure in the galaxy density distribution but no apparent 
secondary structure in the X-ray image. With further optical data, \rxj, Abell~2142 
and MS1455.0+2232 could provide definitive tests of the scenario put forth by 
\citet{ascasibar2006} where an essentially gas-free group perturbs the core ICM,
setting off sloshing whilst providing minimal further evidence of a merger 
at X-ray wavelengths. 

While much of the evidence favoring a merger origin for the relaxed appearing 
sloshing type cold front clusters comes from simulations, two significant 
observations exist which show the plausibility of merger-induced sloshing as a
mechanism for the formation of cold fronts. The XMM observations of
Abell~1644\footnote{Abell~1644 is not part of the cold front sample presented 
in this paper, since its redshift falls below the cutoff of z=0.05.}
\citep{reiprich2004,markevitch2007} and the \chan\, observations of Abell~1201
presented here and in \citet{owers2009a} both reveal sloshing cold fronts and 
also remnant cores which are apparently causing the perturbation. The optical
follow up observations of Abell~1201 showed evidence for localized 
velocity substructure coincident with the remnant core \citep{owers2009a}, 
confirming it is a significant merging substructure, although the latest optical 
observations of Abell~1644 do not reveal significant evidence for localized velocity 
substructure \citep{tustin2001}. Without detailed simulations of these two particular
objects, it is unclear whether they will evolve into relaxed appearing clusters
with cold fronts, like those presented in Section~\ref{relaxed}. Nevertheless,
it is clear that both Abell~1644 and Abell~1201 provide strong empirical evidence 
of merger-induced sloshing as a viable mechanism for producing cold fronts.

Perhaps the most appealing outcome that studies of cold front clusters 
offer is the possibility of using them to distinguish major and minor mergers.
Clearly the clusters in Section~\ref{disturbed} are systems which are undergoing
significant dynamical evolution and are most likely major mergers. However, the 
same conclusion cannot be drawn based on the evidence available for the clusters 
presented in Section~\ref{relaxed}. As discussed above, \citet{ascasibar2006} found 
that only a relatively minor merger was required to produce cold fronts in relaxed
appearing clusters. Future dynamical analyses of \rxj, Abell~2142 and MS1455.0+2232
at optical wavelengths will determine the level of dynamical activity required to
produce a cold front. If it turns out to be true that the appearance of these clusters 
can be explained by minor mergers, then the combination of the existence of a cold front
and X-ray morphology will provide a useful diagnostic for discriminating major and 
minor mergers in clusters.

\section{Summary and Conclusions}\label{summary}

We have presented and described in detail a sample of nine clusters selected 
from the \chan\ archive based purely on the existence of cold fronts 
in their ICM. These clusters are the subject of an ongoing study to determine 
if cold fronts may be used to reliably identify recent cluster-cluster mergers. 

In selecting the cold front sample, particular attention was paid 
to obtaining highly reliable identifications. To this end, each cluster
met the following strict selection criteria: A total \chan\ exposure time 
greater than 40ks, $0.05 < z < 0.3$, a visible surface brightness edge 
in the X-ray image which has a density jump exceeding 1.5 at the lower 
$90\%$ confidence level, the measured gas temperature must be lower 
on the denser side of the front, and the pressure must be continuous 
across the front within the errors. This resulted in the identification of nine 
cold front clusters for which we present temperature maps, X-ray and optical images
and an overview of the multiwavelength evidence for merger activity available in 
the literature. The key conclusions of this study are as follows:

\begin{itemize}
\item the measurement of thermal properties across the fronts was an 
essential diagnostic tool; Abell~2034 and Abell~665 were rejected on this 
basis, since their fronts were better explained as shocks. These 
are very likely merger shocks and if these clusters were to remain in our 
sample, the contamination would not be harmful when considering the aim is to 
use cold fronts as reliable signposts of merger activity. 

\item  There is a dichotomy within the sample of cold front 
clusters: those which harbor clearly disturbed X-ray morphologies, and those 
which appear to have fairly relaxed X-ray morphologies, apart from the existence 
of a cold front.

\item The majority of those clusters with disturbed X-ray morphologies
have significant evidence for merger activity at multiple wavelengths,
while minimal evidence for a merger exists at other wavelengths in the
relaxed appearing clusters.

\item Two of the relaxed appearing clusters (\rxj\ and MS1455.0+2232)
exhibit spiral-like structures in their temperature maps, similar to 
those seen in the simulations of \citet{ascasibar2006}.

\end{itemize}

To conclude, in the case of the cold front clusters with disturbed 
morphologies, it is clear that a cold front is an excellent signpost for
ongoing major merger activity. However, it appears verification that all cold
fronts are signatures of merger activity (major merger or otherwise) relies
critically on the results of follow up observations of the relaxed appearing
cold front clusters. The sample is the subject of extensive follow up 
observations using multi-object spectroscopy at optical wavelengths to
obtain large samples of confirmed cluster members. These observations will be 
used to search for dynamical substructure, constrain merger dynamics and to 
study the effects of the merging environment on the cluster galaxies.

\section{Acknowledgments}
We thank the anonymous referee for useful comments which greatly improved this manuscript.
MSO was supported by an Australian Postgraduate Award, and acknowledges the 
hospitality of the Harvard-Smithsonian Center for Astrophysics where a major 
portion of this study was undertaken. We acknowledge the financial support of 
the Australian Research Council (via its Discovery Project Scheme) throughout 
the course of this work. PEJN was supported by NASA grant NAS8-03060. 

This research has made use of software provided by the Chandra X-ray Center 
(CXC) in the application packages CIAO, ChIPS, and Sherpa and also of data 
obtained from the Chandra archive at the NASA Chandra X-ray center 
(http://cxc.harvard.edu/cda/). This research has made use of the NASA/IPAC 
Extragalactic Database (NED) which is operated by the Jet Propulsion Laboratory,
California Institute of Technology, under contract with the National Aeronautics
and Space Administration. This research used the facilities of the Canadian Astronomy 
Data Centre operated by the National Research Council of Canada with the support 
of the Canadian Space Agency. 

\begin{appendix}
\section{Surface brightness broken power law density  model}\label{sb.model}
For the purpose of quantifying the edges in surface brightness, it is 
assumed that a discontinuity is formed at the interface of gases of differing 
physical properties, occupying two similar concentric spheroids with their 
rotational symmetry axis in the plane of the sky. Thus the curvature of the 
front relative to the line of sight is assumed to be the same as that in the 
plane of the sky. The electron density, $n_e$, is described by the broken power 
law function
\begin{equation}
n_e(r)=\left\{
\begin{array}{ll}
  n_{e,1}\left({r}\over{R_f}\right)^{-\alpha_1}, &  r<R_f,\\
  n_{e,2}\left({r}\over{R_f}\right)^{-\alpha_2}, &  r>R_f,
\end{array}
\right.
\label{sb_ne}
\end{equation}
where $R_f$ is the radius of the discontinuity, $n_{e,1}$ and $n_{e,2}$ are the 
densities of the inner and outer gas at $R_f$, respectively, the elliptical 
radius, $r$, is defined by $r^2=\varpi^2+{\epsilon_\zeta}^2 \zeta^2$, $\zeta$
is the coordinate along our line-of-sight, $\epsilon_\zeta$ determines the axis 
length in that direction and $\varpi$, the elliptical radius in the plane of the
sky, is defined by
\begin{equation}
\varpi^2={(x \cos\theta+y \sin\theta)^2+
\epsilon^2(y \cos\theta-x \sin\theta)^2},
\end{equation}
where $x$ and $y$ are cartesian coordinates with respect to the center of the 
ellipse, $\theta$ is the position angle and $\epsilon$ is the ellipticity. The 
X-ray emission primarily arises from Bremsstrahlung radiation with secondary 
contributions from collisional line excitation radiation. Both these emission 
processes depend on the electron and ion densities in the ICM, and the X-ray 
surface brightness, $S_{\rm X}$, is proportional to the integrated square of the
electron density along the line of sight so that
\begin{equation}
S_{\rm X}={1 \over{4\pi(1+z)^4}}\int F(T,Z){n_e}^2  d\zeta,
\label{sb}
\end{equation} 
where $z$ is the cluster redshift, $n_e$ is the electron density, and $F(T,Z)$ 
gives the dependence of the telescope response to thermal emission from an 
optically thin gas of the temperature, $T$, and abundance, $Z$, (there is also 
some dependence on the column density, $n_H$, and $z$) integrated over
the energy range of interest (0.5-7 keV). The function $F(T,Z)$ also contains
the constant of proportionality relating $n_e$ to the proton density. 
Substituting the relations from Eqns.~\ref{sb_ne} into Eqn.~\ref{sb} and 
integrating, the surface brightness distribution is
\begin{equation}\label{sb.mod}
S_{\rm X}(\varpi)=\left\{
\begin{array}{ll}
A_1 ({\varpi\over{R_F}})^{(1-2\alpha_1)} \beta_{I,1}+
A_2({\varpi\over{R_F}})^{(1-2\alpha_2)}(\beta_2-\beta_{I,2}),&  \varpi<R_f\\ 
A_2({\varpi\over{R_F}})^{(1-2\alpha_2)} \beta_2, &  \varpi>R_f,
\end{array}
\right.
\end{equation}
where $\beta_{I,1}=\beta({1\over{2}}, \alpha_1-{1\over{2}}; 
1-({\varpi\over{R_F}})^2)$, and 
$\beta_{I,2}=\beta({1\over{2}}, \alpha_2-{1\over{2}}; 
1-({\varpi\over{R_F}})^2)$  are incomplete beta functions, and 
$\beta_2=\beta({1\over{2}}, \alpha_2-{1\over{2}})$ is a complete beta function.
For $\alpha_1<0.5$, recursion relations were used to overcome the divergence of 
the incomplete beta function. At $\alpha_1=0.5,-0.5$, the integrals were solved 
analytically. The constants $A_1$ and $A_2$ are
\begin{equation}
A_1={n_{e,1}^2F(T_1,Z_1)R_f \over 4\pi (1+z)^4 \epsilon_{\zeta}},\\
\,A_2={n_{e,2}^2F(T_2,Z_2)R_f \over 4\pi (1+z)^4 \epsilon_{\zeta}},
\end{equation}
and the density jump at the discontinuity can be measured by taking the ratio 
of $A_1$ to $A_2$
\begin{equation}
{n_{e,1}\over n_{e,2}}=\sqrt{{A_1F(T_2,Z_2)\over A_2F(T_1,Z_1)}}
\end{equation}
For temperatures greater than 2\,keV, $F(T,Z)$ is quite insensitive to 
temperature. Thus, the density jump is well approximated by 
${n_{e,1}\over n_{e,2}}\approx\sqrt{{A_1\over A_2}}$.

Note that while the density is discontinuous at the
front, after projection onto the sky, the surface brightness is not.
Thus we refer to the corresponding feature as an ``edge'' in the surface
brightness.  As illustrated in Figure~\ref{sb.profc}, surface brightness does change
abruptly at cold fronts, which can give the misleading impression of
a discontinuity.
\end{appendix}

\clearpage 
\begin{landscape}
  \begin{deluxetable}{ccccccc}
    \tabletypesize{\scriptsize}
    \tablecolumns{6}
    \tablewidth{0pt}
    \tablecaption{Summary of merger evidences and X-ray morphologies for the cold front sample.\label{merger.evidence}}
    \tablehead{\colhead{Cluster} & \colhead{X-ray Morphology}&\colhead{Temperature Map }& \colhead{Optical Substructure\tablenotemark{a}} & \colhead{Radio} & \colhead{Lensing-based}  & \colhead{References}\\
&&Structure&&&mass maps&
    }
    \startdata
    \cutinhead{Disturbed Clusters}
    1ES0657-558& disturbed, remnant core &shock front,  &projected galaxy density,  & halo & substructure  &  1, 3, 10\\
    &&cool remnant core & position plus velocity &&&\\
    Abell~3667 & disturbed&cool mushroom-cap, &projected galaxy density&relic, NATs &substructure& 9, 17, 18\\
    &                       &hot patches & position plus velocity &           &\\
    Abell~1201& remnant core, elliptical, &cool finger joining & projected galaxy density,& \nodata&\nodata & 16\\
    &cool core&cold front and cool core &position plus velocity&&&\\
    Abell~2069 & elliptical, remnant core&cool remnant core,  & projected galaxy density& \nodata&\nodata& 6\\
    &&hot patches&&&&\\ 
    Abell~1758N& disturbed, bimodal  &cool remnant core & projected galaxy density & halo, NATs&substructure& 4, 8, 12, 14\\
    Abell~2163 &disturbed, remnant core& cool remnant core, &projected galaxy density& halo, NATs&substructure&  5, 11, 15, 19\\
    &                       & hot patches & position plus velocity &           &\\
    
    \cutinhead{Relaxed Appearing Clusters}
    RXJ1720.1+2638 &relaxed, cool core&cool spiral, cool core, &projected galaxy density,  &\nodata&possible substructure &4, 15\\
    &&hot patches &possible position plus velocity &&&\\
    MS1455.0+2232& relaxed, cool core, &cool spiral, cool core &\nodata&\nodata& elliptical distribution  & 4\\
    &mildly elliptical& &&&&\\
    Abell~2142 &elliptical, cool core&cool core, cool finger, &possible position plus velocity& NATs, & substructure & 2, 7, 13, 14\\
    &&hot patches &&possible halo&&\\
    \enddata
    \tablenotetext{a}{Position plus velocity substructure refers to substructure found in tests which combine spatial and velocity information. 
      Projected galaxy density substructure refers to bimodality or multimodality in projected galaxy density maps.}
    \tablerefs{
      (1) \cite{barrena2002};
      (2) \cite{bliton1998};
      (3) \cite{clowe2006};
      (4) \cite{dahle2002};
      (5) \cite{feretti2001};
      (6) \cite{gioia1982};
      (7) \cite{giovannini1999};
      (8) \cite{giovannini2006};
      (9) \cite{joffre2000};
      (10) \cite{liang2000};
      (11) \cite{maurogordato2008};
      (12) \cite{odea1985};
      (13) \cite{oegerle1995};
      (14) \cite{okabe2008};
      (15) \cite{owers2008};
      (16) \cite{owers2009a};
      (17) \cite{owers2009b};
      (18) \cite{rottgering1997};
      (19) \cite{squires1997}
    }
  \end{deluxetable}
  \clearpage
\end{landscape}

\label{lastpage}

\begin{thebibliography}{96}
\expandafter\ifx\csname natexlab\endcsname\relax\def\natexlab#1{#1}\fi

\bibitem[{{Abell} {et~al.}(1989){Abell}, {Corwin}, \& {Olowin}}]{abell1989}
{Abell}, G.~O., {Corwin}, Jr., H.~G., \& {Olowin}, R.~P. 1989, \apjs, 70, 1

\bibitem[{{Allen} \& {Fabian}(1998)}]{allen1998}
{Allen}, S.~W., \& {Fabian}, A.~C. 1998, \mnras, 297, L57

\bibitem[{{Allen} {et~al.}(1996){Allen}, {Fabian}, {Edge}, {Bautz}, {Furuzawa},
  \& {Tawara}}]{allen1996}
{Allen}, S.~W., {Fabian}, A.~C., {Edge}, A.~C., {Bautz}, M.~W., {Furuzawa}, A.,
  \& {Tawara}, Y. 1996, \mnras, 283, 263

\bibitem[{{Anders} \& {Grevesse}(1989)}]{anders1989}
{Anders}, E., \& {Grevesse}, N. 1989, \gca, 53, 197

\bibitem[{{Arnaud}(1996)}]{1996ASPC..101...17A}
{Arnaud}, K.~A. 1996, in Astronomical Society of the Pacific Conference Series,
  Vol. 101, Astronomical Data Analysis Software and Systems V, ed. G.~H.
  {Jacoby} \& J.~{Barnes}, 17--+

\bibitem[{{Arnaud} {et~al.}(1992){Arnaud}, {Hughes}, {Forman}, {Jones},
  {Lachieze-Rey}, {Yamashita}, \& {Hatsukade}}]{arnaud1992}
{Arnaud}, M., {Hughes}, J.~P., {Forman}, W., {Jones}, C., {Lachieze-Rey}, M.,
  {Yamashita}, K., \& {Hatsukade}, I. 1992, \apj, 390, 345

\bibitem[{{Ascasibar} \& {Markevitch}(2006)}]{ascasibar2006}
{Ascasibar}, Y., \& {Markevitch}, M. 2006, \apj, 650, 102

\bibitem[{{Barrena} {et~al.}(2002){Barrena}, {Biviano}, {Ramella}, {Falco}, \&
  {Seitz}}]{barrena2002}
{Barrena}, R., {Biviano}, A., {Ramella}, M., {Falco}, E.~E., \& {Seitz}, S.
  2002, \aap, 386, 816

\bibitem[{{Bialek} {et~al.}(2002){Bialek}, {Evrard}, \& {Mohr}}]{bialek2002}
{Bialek}, J.~J., {Evrard}, A.~E., \& {Mohr}, J.~J. 2002, \apjl, 578, L9

\bibitem[{{Bliton} {et~al.}(1998){Bliton}, {Rizza}, {Burns}, {Owen}, \&
  {Ledlow}}]{bliton1998}
{Bliton}, M., {Rizza}, E., {Burns}, J.~O., {Owen}, F.~N., \& {Ledlow}, M.~J.
  1998, \mnras, 301, 609

\bibitem[{{B{\"o}hringer} {et~al.}(2000){B{\"o}hringer}, {Voges}, {Huchra},
  {McLean}, {Giacconi}, {Rosati}, {Burg}, {Mader}, {Schuecker}, {Simi{\c c}},
  {Komossa}, {Reiprich}, {Retzlaff}, \& {Tr{\"u}mper}}]{bohringer2000}
{B{\"o}hringer}, H., {Voges}, W., {Huchra}, J.~P., {McLean}, B., {Giacconi},
  R., {Rosati}, P., {Burg}, R., {Mader}, J., {Schuecker}, P., {Simi{\c c}}, D.,
  {Komossa}, S., {Reiprich}, T.~H., {Retzlaff}, J., \& {Tr{\"u}mper}, J. 2000,
  \apjs, 129, 435

\bibitem[{{Brada{\v c}} {et~al.}(2006){Brada{\v c}}, {Clowe}, {Gonzalez},
  {Marshall}, {Forman}, {Jones}, {Markevitch}, {Randall}, {Schrabback}, \&
  {Zaritsky}}]{bradac2006}
{Brada{\v c}}, M., {Clowe}, D., {Gonzalez}, A.~H., {Marshall}, P., {Forman},
  W., {Jones}, C., {Markevitch}, M., {Randall}, S., {Schrabback}, T., \&
  {Zaritsky}, D. 2006, \apj, 652, 937

\bibitem[{{Briel} {et~al.}(2004){Briel}, {Finoguenov}, \& {Henry}}]{briel2004}
{Briel}, U.~G., {Finoguenov}, A., \& {Henry}, J.~P. 2004, \aap, 426, 1

\bibitem[{{Buote} \& {Tsai}(1996)}]{buote1996}
{Buote}, D.~A., \& {Tsai}, J.~C. 1996, \apj, 458, 27

\bibitem[{{Churazov} {et~al.}(2003){Churazov}, {Forman}, {Jones}, \&
  {B{\"o}hringer}}]{churazov2003}
{Churazov}, E., {Forman}, W., {Jones}, C., \& {B{\"o}hringer}, H. 2003, \apj,
  590, 225

\bibitem[{{Churazov} \& {Inogamov}(2004)}]{churazov2004}
{Churazov}, E., \& {Inogamov}, N. 2004, \mnras, 350, L52

\bibitem[{{Clowe} {et~al.}(2006){Clowe}, {Brada{\v c}}, {Gonzalez},
  {Markevitch}, {Randall}, {Jones}, \& {Zaritsky}}]{clowe2006}
{Clowe}, D., {Brada{\v c}}, M., {Gonzalez}, A.~H., {Markevitch}, M., {Randall},
  S.~W., {Jones}, C., \& {Zaritsky}, D. 2006, \apjl, 648, L109

\bibitem[{{Cole} {et~al.}(2005){Cole}, {Percival}, {Peacock}, {Norberg},
  {Baugh}, {Frenk}, {Baldry}, {Bland-Hawthorn}, {Bridges}, {Cannon}, {Colless},
  {Collins}, {Couch}, {Cross}, {Dalton}, {Eke}, {De Propris}, {Driver},
  {Efstathiou}, {Ellis}, {Glazebrook}, {Jackson}, {Jenkins}, {Lahav}, {Lewis},
  {Lumsden}, {Maddox}, {Madgwick}, {Peterson}, {Sutherland}, \&
  {Taylor}}]{cole2005}
{Cole}, S., {Percival}, W.~J., {Peacock}, J.~A., {Norberg}, P., {Baugh}, C.~M.,
  {Frenk}, C.~S., {Baldry}, I., {Bland-Hawthorn}, J., {Bridges}, T., {Cannon},
  R., {Colless}, M., {Collins}, C., {Couch}, W., {Cross}, N.~J.~G., {Dalton},
  G., {Eke}, V.~R., {De Propris}, R., {Driver}, S.~P., {Efstathiou}, G.,
  {Ellis}, R.~S., {Glazebrook}, K., {Jackson}, C., {Jenkins}, A., {Lahav}, O.,
  {Lewis}, I., {Lumsden}, S., {Maddox}, S., {Madgwick}, D., {Peterson}, B.~A.,
  {Sutherland}, W., \& {Taylor}, K. 2005, \mnras, 362, 505

\bibitem[{{Dahle} {et~al.}(2002){Dahle}, {Kaiser}, {Irgens}, {Lilje}, \&
  {Maddox}}]{dahle2002}
{Dahle}, H., {Kaiser}, N., {Irgens}, R.~J., {Lilje}, P.~B., \& {Maddox}, S.~J.
  2002, \apjs, 139, 313

\bibitem[{{David} \& {Kempner}(2004)}]{david2004}
{David}, L.~P., \& {Kempner}, J. 2004, \apj, 613, 831

\bibitem[{{Dickey} \& {Lockman}(1990)}]{1990ARA&A..28..215D}
{Dickey}, J.~M., \& {Lockman}, F.~J. 1990, \araa, 28, 215

\bibitem[{{Ebeling} {et~al.}(1996){Ebeling}, {Voges}, {Bohringer}, {Edge},
  {Huchra}, \& {Briel}}]{ebeling1996}
{Ebeling}, H., {Voges}, W., {Bohringer}, H., {Edge}, A.~C., {Huchra}, J.~P., \&
  {Briel}, U.~G. 1996, \mnras, 281, 799

\bibitem[{{Edge} {et~al.}(2003){Edge}, {Smith}, {Sand}, {Treu}, {Ebeling},
  {Allen}, \& {van Dokkum}}]{edge2003}
{Edge}, A.~C., {Smith}, G.~P., {Sand}, D.~J., {Treu}, T., {Ebeling}, H.,
  {Allen}, S.~W., \& {van Dokkum}, P.~G. 2003, \apjl, 599, L69

\bibitem[{{Einasto} {et~al.}(1997){Einasto}, {Tago}, {Jaaniste}, {Einasto}, \&
  {Andernach}}]{einasto1997}
{Einasto}, M., {Tago}, E., {Jaaniste}, J., {Einasto}, J., \& {Andernach}, H.
  1997, \aaps, 123, 119

\bibitem[{{Elbaz} {et~al.}(1995){Elbaz}, {Arnaud}, \& {Boehringer}}]{elbaz1995}
{Elbaz}, D., {Arnaud}, M., \& {Boehringer}, H. 1995, \aap, 293, 337

\bibitem[{{Elvis} {et~al.}(1992){Elvis}, {Plummer}, {Schachter}, \&
  {Fabbiano}}]{elvis1992}
{Elvis}, M., {Plummer}, D., {Schachter}, J., \& {Fabbiano}, G. 1992, \apjs, 80,
  257

\bibitem[{{Feretti} {et~al.}(2001){Feretti}, {Fusco-Femiano}, {Giovannini}, \&
  {Govoni}}]{feretti2001}
{Feretti}, L., {Fusco-Femiano}, R., {Giovannini}, G., \& {Govoni}, F. 2001,
  \aap, 373, 106

\bibitem[{{Feretti} {et~al.}(2004){Feretti}, {Orr{\`u}}, {Brunetti},
  {Giovannini}, {Kassim}, \& {Setti}}]{feretti2004}
{Feretti}, L., {Orr{\`u}}, E., {Brunetti}, G., {Giovannini}, G., {Kassim}, N.,
  \& {Setti}, G. 2004, \aap, 423, 111

\bibitem[{{Fujita} {et~al.}(2004){Fujita}, {Sarazin}, {Reiprich}, {Andernach},
  {Ehle}, {Murgia}, {Rudnick}, \& {Slee}}]{fujita2004}
{Fujita}, Y., {Sarazin}, C.~L., {Reiprich}, T.~H., {Andernach}, H., {Ehle}, M.,
  {Murgia}, M., {Rudnick}, L., \& {Slee}, O.~B. 2004, \apj, 616, 157

\bibitem[{{Gioia} {et~al.}(1982){Gioia}, {Maccacaro}, {Geller}, {Huchra},
  {Stocke}, \& {Steiner}}]{gioia1982}
{Gioia}, I.~M., {Maccacaro}, T., {Geller}, M.~J., {Huchra}, J.~P., {Stocke},
  J., \& {Steiner}, J.~E. 1982, \apjl, 255, L17

\bibitem[{{Gioia} {et~al.}(1990){Gioia}, {Maccacaro}, {Schild}, {Wolter},
  {Stocke}, {Morris}, \& {Henry}}]{gioia1990}
{Gioia}, I.~M., {Maccacaro}, T., {Schild}, R.~E., {Wolter}, A., {Stocke},
  J.~T., {Morris}, S.~L., \& {Henry}, J.~P. 1990, \apjs, 72, 567

\bibitem[{{Giovannini} {et~al.}(2006){Giovannini}, {Feretti}, {Govoni},
  {Murgia}, \& {Pizzo}}]{giovannini2006}
{Giovannini}, G., {Feretti}, L., {Govoni}, F., {Murgia}, M., \& {Pizzo}, R.
  2006, Astronomische Nachrichten, 327, 563

\bibitem[{{Giovannini} {et~al.}(1999){Giovannini}, {Tordi}, \&
  {Feretti}}]{giovannini1999}
{Giovannini}, G., {Tordi}, M., \& {Feretti}, L. 1999, New Astronomy, 4, 141

\bibitem[{{Govoni} {et~al.}(2004){Govoni}, {Markevitch}, {Vikhlinin},
  {VanSpeybroeck}, {Feretti}, \& {Giovannini}}]{govoni2004}
{Govoni}, F., {Markevitch}, M., {Vikhlinin}, A., {VanSpeybroeck}, L.,
  {Feretti}, L., \& {Giovannini}, G. 2004, \apj, 605, 695

\bibitem[{{Heinz} {et~al.}(2003){Heinz}, {Churazov}, {Forman}, {Jones}, \&
  {Briel}}]{heinz2003}
{Heinz}, S., {Churazov}, E., {Forman}, W., {Jones}, C., \& {Briel}, U.~G. 2003,
  \mnras, 346, 13

\bibitem[{{Henry} \& {Briel}(1996)}]{henry1996}
{Henry}, J.~P., \& {Briel}, U.~G. 1996, \apj, 472, 137

\bibitem[{{Hernquist} {et~al.}(1996){Hernquist}, {Katz}, {Weinberg}, \&
  {Miralda-Escud{\'e}}}]{hernquist1996}
{Hernquist}, L., {Katz}, N., {Weinberg}, D.~H., \& {Miralda-Escud{\'e}}, J.
  1996, \apjl, 457, L51+

\bibitem[{{Holzapfel} {et~al.}(1997){Holzapfel}, {Arnaud}, {Ade}, {Church},
  {Fischer}, {Mauskopf}, {Rephaeli}, {Wilbanks}, \& {Lange}}]{holzapfel1997}
{Holzapfel}, W.~L., {Arnaud}, M., {Ade}, P.~A.~R., {Church}, S.~E., {Fischer},
  M.~L., {Mauskopf}, P.~D., {Rephaeli}, Y., {Wilbanks}, T.~M., \& {Lange},
  A.~E. 1997, \apj, 480, 449

\bibitem[{{Jeltema} {et~al.}(2005){Jeltema}, {Canizares}, {Bautz}, \&
  {Buote}}]{jeltema2005}
{Jeltema}, T.~E., {Canizares}, C.~R., {Bautz}, M.~W., \& {Buote}, D.~A. 2005,
  \apj, 624, 606

\bibitem[{{Joffre} {et~al.}(2000){Joffre}, {Fischer}, {Frieman}, {McKay},
  {Mohr}, {Nichol}, {Johnston}, {Sheldon}, \& {Bernstein}}]{joffre2000}
{Joffre}, M., {Fischer}, P., {Frieman}, J., {McKay}, T., {Mohr}, J.~J.,
  {Nichol}, R.~C., {Johnston}, D., {Sheldon}, E., \& {Bernstein}, G. 2000,
  \apjl, 534, L131

\bibitem[{{Kaastra}(1992)}]{kaastra1992}
{Kaastra}, J.~S. 1992, An X-Ray Spectral Code for Optically Thin Plasmas
  (Internal SRON-Leiden Report, updated version 2.0)

\bibitem[{{Knopp} {et~al.}(1996){Knopp}, {Henry}, \& {Briel}}]{knopp1996}
{Knopp}, G.~P., {Henry}, J.~P., \& {Briel}, U.~G. 1996, \apj, 472, 125

\bibitem[{{Lacey} \& {Cole}(1993)}]{lacey1993}
{Lacey}, C., \& {Cole}, S. 1993, \mnras, 262, 627

\bibitem[{{Liang} {et~al.}(2000){Liang}, {Hunstead}, {Birkinshaw}, \&
  {Andreani}}]{liang2000}
{Liang}, H., {Hunstead}, R.~W., {Birkinshaw}, M., \& {Andreani}, P. 2000, \apj,
  544, 686

\bibitem[{{Liedahl} {et~al.}(1995){Liedahl}, {Osterheld}, \&
  {Goldstein}}]{1995ApJ...438L.115L}
{Liedahl}, D.~A., {Osterheld}, A.~L., \& {Goldstein}, W.~H. 1995, \apjl, 438,
  L115

\bibitem[{{Lyutikov}(2006)}]{lyutikov2006}
{Lyutikov}, M. 2006, \mnras, 373, 73

\bibitem[{{Machacek} {et~al.}(2005){Machacek}, {Dosaj}, {Forman}, {Jones},
  {Markevitch}, {Vikhlinin}, {Warmflash}, \& {Kraft}}]{machacek2005}
{Machacek}, M., {Dosaj}, A., {Forman}, W., {Jones}, C., {Markevitch}, M.,
  {Vikhlinin}, A., {Warmflash}, A., \& {Kraft}, R. 2005, \apj, 621, 663

\bibitem[{{Machacek} {et~al.}(2006){Machacek}, {Jones}, {Forman}, \&
  {Nulsen}}]{machacek2006}
{Machacek}, M., {Jones}, C., {Forman}, W.~R., \& {Nulsen}, P. 2006, \apj, 644,
  155

\bibitem[{{Markevitch}(1996)}]{markevitch1996}
{Markevitch}, M. 1996, \apjl, 465, L1+

\bibitem[{{Markevitch} {et~al.}(2003{\natexlab{a}}){Markevitch}, {Bautz},
  {Biller}, {Butt}, {Edgar}, {Gaetz}, {Garmire}, {Grant}, {Green}, {Juda},
  {Plucinsky}, {Schwartz}, {Smith}, {Vikhlinin}, {Virani}, {Wargelin}, \&
  {Wolk}}]{markevitch2003}
{Markevitch}, M., {Bautz}, M.~W., {Biller}, B., {Butt}, Y., {Edgar}, R.,
  {Gaetz}, T., {Garmire}, G., {Grant}, C.~E., {Green}, P., {Juda}, M.,
  {Plucinsky}, P.~P., {Schwartz}, D., {Smith}, R., {Vikhlinin}, A., {Virani},
  S., {Wargelin}, B.~J., \& {Wolk}, S. 2003{\natexlab{a}}, \apj, 583, 70

\bibitem[{{Markevitch} {et~al.}(1998){Markevitch}, {Forman}, {Sarazin}, \&
  {Vikhlinin}}]{markevitch1998}
{Markevitch}, M., {Forman}, W.~R., {Sarazin}, C.~L., \& {Vikhlinin}, A. 1998,
  \apj, 503, 77

\bibitem[{{Markevitch} {et~al.}(2002){Markevitch}, {Gonzalez}, {David},
  {Vikhlinin}, {Murray}, {Forman}, {Jones}, \& {Tucker}}]{markevitch2002}
{Markevitch}, M., {Gonzalez}, A.~H., {David}, L., {Vikhlinin}, A., {Murray},
  S., {Forman}, W., {Jones}, C., \& {Tucker}, W. 2002, \apjl, 567, L27

\bibitem[{{Markevitch} {et~al.}(2000){Markevitch}, {Ponman}, {Nulsen}, {Bautz},
  {Burke}, {David}, {Davis}, {Donnelly}, {Forman}, {Jones}, {Kaastra},
  {Kellogg}, {Kim}, {Kolodziejczak}, {Mazzotta}, {Pagliaro}, {Patel}, {Van
  Speybroeck}, {Vikhlinin}, {Vrtilek}, {Wise}, \& {Zhao}}]{markevitch2000}
{Markevitch}, M., {Ponman}, T.~J., {Nulsen}, P.~E.~J., {Bautz}, M.~W., {Burke},
  D.~J., {David}, L.~P., {Davis}, D., {Donnelly}, R.~H., {Forman}, W.~R.,
  {Jones}, C., {Kaastra}, J., {Kellogg}, E., {Kim}, D.-W., {Kolodziejczak}, J.,
  {Mazzotta}, P., {Pagliaro}, A., {Patel}, S., {Van Speybroeck}, L.,
  {Vikhlinin}, A., {Vrtilek}, J., {Wise}, M., \& {Zhao}, P. 2000, \apj, 541,
  542

\bibitem[{{Markevitch} {et~al.}(1999){Markevitch}, {Sarazin}, \&
  {Vikhlinin}}]{markevitch1999}
{Markevitch}, M., {Sarazin}, C.~L., \& {Vikhlinin}, A. 1999, \apj, 521, 526

\bibitem[{{Markevitch} \& {Vikhlinin}(2001)}]{markevitch2001a}
{Markevitch}, M., \& {Vikhlinin}, A. 2001, \apj, 563, 95

\bibitem[{{Markevitch} \& {Vikhlinin}(2007)}]{markevitch2007}
---. 2007, \physrep, 443, 1

\bibitem[{{Markevitch} {et~al.}(2003{\natexlab{b}}){Markevitch}, {Vikhlinin},
  \& {Forman}}]{markevitch2003asp}
{Markevitch}, M., {Vikhlinin}, A., \& {Forman}, W.~R. 2003{\natexlab{b}}, in
  Astronomical Society of the Pacific Conference Series, Vol. 301, Astronomical
  Society of the Pacific Conference Series, ed. S.~{Bowyer} \& C.-Y. {Hwang},
  37--+

\bibitem[{{Markevitch} {et~al.}(2001){Markevitch}, {Vikhlinin}, \&
  {Mazzotta}}]{markevitch2001}
{Markevitch}, M., {Vikhlinin}, A., \& {Mazzotta}, P. 2001, \apjl, 562, L153

\bibitem[{{Markevitch} {et~al.}(1994){Markevitch}, {Yamashita}, {Furuzawa}, \&
  {Tawara}}]{markevitch1994}
{Markevitch}, M., {Yamashita}, K., {Furuzawa}, A., \& {Tawara}, Y. 1994, \apjl,
  436, L71

\bibitem[{{Mastropietro} \& {Burkert}(2008)}]{mastropietro2008}
{Mastropietro}, C., \& {Burkert}, A. 2008, \mnras, 389, 967

\bibitem[{{Mathis} {et~al.}(2005){Mathis}, {Lavaux}, {Diego}, \&
  {Silk}}]{mathis2005}
{Mathis}, H., {Lavaux}, G., {Diego}, J.~M., \& {Silk}, J. 2005, \mnras, 357,
  801

\bibitem[{{Maughan} {et~al.}(2008){Maughan}, {Jones}, {Forman}, \& {Van
  Speybroeck}}]{maughan2008}
{Maughan}, B.~J., {Jones}, C., {Forman}, W., \& {Van Speybroeck}, L. 2008,
  \apjs, 174, 117

\bibitem[{{Maurogordato} {et~al.}(2008){Maurogordato}, {Cappi}, {Ferrari},
  {Benoist}, {Mars}, {Soucail}, {Arnaud}, {Pratt}, {Bourdin}, \&
  {Sauvageot}}]{maurogordato2008}
{Maurogordato}, S., {Cappi}, A., {Ferrari}, C., {Benoist}, C., {Mars}, G.,
  {Soucail}, G., {Arnaud}, M., {Pratt}, G.~W., {Bourdin}, H., \& {Sauvageot},
  J.-L. 2008, \aap, 481, 593

\bibitem[{{Mazzotta} {et~al.}(2002){Mazzotta}, {Fusco-Femiano}, \&
  {Vikhlinin}}]{mazzotta2002}
{Mazzotta}, P., {Fusco-Femiano}, R., \& {Vikhlinin}, A. 2002, \apjl, 569, L31

\bibitem[{{Mazzotta} \& {Giacintucci}(2008)}]{mazzotta2008}
{Mazzotta}, P., \& {Giacintucci}, S. 2008, \apjl, 675, L9

\bibitem[{{Mazzotta} {et~al.}(2001){Mazzotta}, {Markevitch}, {Vikhlinin},
  {Forman}, {David}, \& {VanSpeybroeck}}]{mazzotta2001}
{Mazzotta}, P., {Markevitch}, M., {Vikhlinin}, A., {Forman}, W.~R., {David},
  L.~P., \& {VanSpeybroeck}, L. 2001, \apj, 555, 205

\bibitem[{{Mewe} {et~al.}(1985){Mewe}, {Gronenschild}, {van den Oord}}]{mewe1985}
{Mewe}, R., {Gronenschild}, E.~H.~B.~M., \& {van den Oord}, G.~H.~J. 1985, \aaps, 62, 197

\bibitem[{{Mewe} {et~al.}(1986){Mewe}, {Lemen}, {van den Oord}}]{mewe1986}
{Mewe}, R., {Lemen}, J.~R., \& {van den Oord}, G.~H.~J. 1986, \aaps, 65, 511

\bibitem[{{Million} \& {Allen}(2008)}]{million2008}
{Million}, E.~T., \& {Allen}, S.~W. 2008, arXiv:astro-ph/0108476

\bibitem[{{Milosavljevi{\'c}} {et~al.}(2007){Milosavljevi{\'c}}, {Koda},
  {Nagai}, {Nakar}, \& {Shapiro}}]{milosav2007}
{Milosavljevi{\'c}}, M., {Koda}, J., {Nagai}, D., {Nakar}, E., \& {Shapiro},
  P.~R. 2007, \apjl, 661, L131

\bibitem[{{Nagai} \& {Kravtsov}(2003)}]{nagai2003}
{Nagai}, D., \& {Kravtsov}, A.~V. 2003, \apj, 587, 514

\bibitem[{{Nulsen} {et~al.}(2005){Nulsen}, {McNamara}, {Wise}, \&
  {David}}]{nulsen2005}
{Nulsen}, P.~E.~J., {McNamara}, B.~R., {Wise}, M.~W., \& {David}, L.~P. 2005,
  \apj, 628, 629

\bibitem[{{O'dea} \& {Owen}(1985)}]{odea1985}
{O'dea}, C.~P., \& {Owen}, F.~N. 1985, \aj, 90, 927

\bibitem[{{Oegerle} {et~al.}(1995){Oegerle}, {Hill}, \&
  {Fitchett}}]{oegerle1995}
{Oegerle}, W.~R., {Hill}, J.~M., \& {Fitchett}, M.~J. 1995, \aj, 110, 32

\bibitem[{{Okabe} \& {Umetsu}(2008)}]{okabe2008}
{Okabe}, N., \& {Umetsu}, K. 2008, \pasj, 60, 345

\bibitem[{{Onuora} {et~al.}(2003){Onuora}, {Kay}, \& {Thomas}}]{onuora2003}
{Onuora}, L.~I., {Kay}, S.~T., \& {Thomas}, P.~A. 2003, \mnras, 341, 1246

\bibitem[{{Owers}(2008)}]{owers2008}
{Owers}, M.~S. 2008, PhD thesis, University of New South Wales,
  http://handle.unsw.edu.au/1959.4/38964

\bibitem[{{Owers} {et~al.}(2009{\natexlab{a}}){Owers}, {Couch}, \&
  {Nulsen}}]{owers2009b}
{Owers}, M.~S., {Couch}, W.~J., \& {Nulsen}, P.~E.~J. 2009{\natexlab{a}}, \apj,
  693, 901

\bibitem[{{Owers} {et~al.}(2009{\natexlab{b}}){Owers}, {Nulsen}, {Couch},
  {Markevitch}, \& {Poole}}]{owers2009a}
{Owers}, M.~S., {Nulsen}, P.~E.~J., {Couch}, W.~J., {Markevitch}, M., \&
  {Poole}, G.~B. 2009{\natexlab{b}}, \apj, 692, 702

\bibitem[{{Peebles}(1993)}]{peebles1993}
{Peebles}, P.~J.~E. 1993, {Principles of physical cosmology} (Princeton Series
  in Physics, Princeton, NJ: Princeton University Press, |c1993)

\bibitem[{{Poole} {et~al.}(2006){Poole}, {Fardal}, {Babul}, {McCarthy},
  {Quinn}, \& {Wadsley}}]{poole2006}
{Poole}, G.~B., {Fardal}, M.~A., {Babul}, A., {McCarthy}, I.~G., {Quinn}, T.,
  \& {Wadsley}, J. 2006, \mnras, 373, 881

\bibitem[{{Press} \& {Schechter}(1974)}]{press1974}
{Press}, W.~H., \& {Schechter}, P. 1974, \apj, 187, 425

\bibitem[{{Proust} {et~al.}(1988){Proust}, {Mazure}, {Sodre}, {Capelato}, \&
  {Lund}}]{proust1988}
{Proust}, D., {Mazure}, A., {Sodre}, L., {Capelato}, H., \& {Lund}, G. 1988,
  \aaps, 72, 415

\bibitem[{{Reiprich} {et~al.}(2004){Reiprich}, {Sarazin}, {Kempner}, \&
  {Tittley}}]{reiprich2004}
{Reiprich}, T.~H., {Sarazin}, C.~L., {Kempner}, J.~C., \& {Tittley}, E. 2004,
  \apj, 608, 179

\bibitem[{{Roettiger} {et~al.}(1999){Roettiger}, {Burns}, \&
  {Stone}}]{roettiger1999}
{Roettiger}, K., {Burns}, J.~O., \& {Stone}, J.~M. 1999, \apj, 518, 603

\bibitem[{{Rottgering} {et~al.}(1997){Rottgering}, {Wieringa}, {Hunstead}, \&
  {Ekers}}]{rottgering1997}
{Rottgering}, H.~J.~A., {Wieringa}, M.~H., {Hunstead}, R.~W., \& {Ekers}, R.~D.
  1997, \mnras, 290, 577

\bibitem[{{Smail} {et~al.}(1995){Smail}, {Ellis}, {Fitchett}, \&
  {Edge}}]{smail1995}
{Smail}, I., {Ellis}, R.~S., {Fitchett}, M.~J., \& {Edge}, A.~C. 1995, \mnras,
  273, 277

\bibitem[{{Smoot} {et~al.}(1992){Smoot}, {Bennett}, {Kogut}, {Wright}, {Aymon},
  {Boggess}, {Cheng}, {de Amici}, {Gulkis}, {Hauser}, {Hinshaw}, {Jackson},
  {Janssen}, {Kaita}, {Kelsall}, {Keegstra}, {Lineweaver}, {Loewenstein},
  {Lubin}, {Mather}, {Meyer}, {Moseley}, {Murdock}, {Rokke}, {Silverberg},
  {Tenorio}, {Weiss}, \& {Wilkinson}}]{smoot1992}
{Smoot}, G.~F., {Bennett}, C.~L., {Kogut}, A., {Wright}, E.~L., {Aymon}, J.,
  {Boggess}, N.~W., {Cheng}, E.~S., {de Amici}, G., {Gulkis}, S., {Hauser},
  M.~G., {Hinshaw}, G., {Jackson}, P.~D., {Janssen}, M., {Kaita}, E.,
  {Kelsall}, T., {Keegstra}, P., {Lineweaver}, C., {Loewenstein}, K., {Lubin},
  P., {Mather}, J., {Meyer}, S.~S., {Moseley}, S.~H., {Murdock}, T., {Rokke},
  L., {Silverberg}, R.~F., {Tenorio}, L., {Weiss}, R., \& {Wilkinson}, D.~T.
  1992, \apjl, 396, L1

\bibitem[{{Sodre} {et~al.}(1992){Sodre}, {Capelato}, {Steiner}, {Proust}, \&
  {Mazure}}]{sodre1992}
{Sodre}, L.~J., {Capelato}, H.~V., {Steiner}, J.~E., {Proust}, D., \& {Mazure},
  A. 1992, \mnras, 259, 233

\bibitem[{{Springel} \& {Farrar}(2007)}]{springel2007}
{Springel}, V., \& {Farrar}, G.~R. 2007, \mnras, 380, 911

\bibitem[{{Springel} {et~al.}(2006){Springel}, {Frenk}, \&
  {White}}]{springel2006}
{Springel}, V., {Frenk}, C.~S., \& {White}, S.~D.~M. 2006, \nat, 440, 1137

\bibitem[{{Squires} {et~al.}(1997){Squires}, {Neumann}, {Kaiser}, {Arnaud},
  {Babul}, {Boehringer}, {Fahlman}, \& {Woods}}]{squires1997}
{Squires}, G., {Neumann}, D.~M., {Kaiser}, N., {Arnaud}, M., {Babul}, A.,
  {Boehringer}, H., {Fahlman}, G., \& {Woods}, D. 1997, \apj, 482, 648

\bibitem[{{Struble} \& {Rood}(1999)}]{struble1999}
{Struble}, M.~F., \& {Rood}, H.~J. 1999, \apjs, 125, 35

\bibitem[{{Tittley} \& {Henriksen}(2005)}]{tittley2005}
{Tittley}, E.~R., \& {Henriksen}, M. 2005, \apj, 618, 227

\bibitem[{{Tucker} {et~al.}(1998){Tucker}, {Blanco}, {Rappoport}, {David},
  {Fabricant}, {Falco}, {Forman}, {Dressler}, \& {Ramella}}]{tucker1998}
{Tucker}, W., {Blanco}, P., {Rappoport}, S., {David}, L., {Fabricant}, D.,
  {Falco}, E.~E., {Forman}, W., {Dressler}, A., \& {Ramella}, M. 1998, \apjl,
  496, L5+

\bibitem[{{Tustin} {et~al.}(2001){Tustin}, {Geller}, {Kenyon}, \&
  {Diaferio}}]{tustin2001}
{Tustin}, A.~W., {Geller}, M.~J., {Kenyon}, S.~J., \& {Diaferio}, A. 2001, \aj,
  122, 1289

\bibitem[{{Vikhlinin} {et~al.}(2001{\natexlab{a}}){Vikhlinin}, {Markevitch}, \&
  {Murray}}]{vikhlinin2001b}
{Vikhlinin}, A., {Markevitch}, M., \& {Murray}, S.~S. 2001{\natexlab{a}}, \apj,
  551, 160

\bibitem[{{Vikhlinin} {et~al.}(2001{\natexlab{b}}){Vikhlinin}, {Markevitch}, \&
  {Murray}}]{vikhlinin2001a}
---. 2001{\natexlab{b}}, \apjl, 549, L47

\bibitem[{{Weisskopf} {et~al.}(2002){Weisskopf}, {Brinkman}, {Canizares},
  {Garmire}, {Murray}, \& {Van Speybroeck}}]{weisskopf2002}
{Weisskopf}, M.~C., {Brinkman}, B., {Canizares}, C., {Garmire}, G., {Murray},
  S., \& {Van Speybroeck}, L.~P. 2002, \pasp, 114, 1

\bibitem[{{White} {et~al.}(1994){White}, {Day}, {Hatsukade}, \&
  {Hughes}}]{white1994}
{White}, III, R.~E., {Day}, C.~S.~R., {Hatsukade}, I., \& {Hughes}, J.~P. 1994,
  \apj, 433, 583

\bibitem[{{Xiang} {et~al.}(2007){Xiang}, {Churazov}, {Dolag}, {Springel}, \&
  {Vikhlinin}}]{xiang2007}
{Xiang}, F., {Churazov}, E., {Dolag}, K., {Springel}, V., \& {Vikhlinin}, A.
  2007, \mnras, 379, 1325

\end{thebibliography}
\end{document}